\documentclass{jltp}
\usepackage[latin9]{inputenc}
\usepackage[T1]{fontenc}
\usepackage[finnish,english]{babel}
\usepackage{amsmath}
\usepackage{graphicx}
\usepackage{color}
\usepackage{pstricks}
\usepackage{pst-node}
\usepackage{subfigure}
\usepackage{amssymb}
\usepackage{cite}

\makeatletter
\renewcommand{\@cite}[2]{%
  \leavevmode\cite@adjust\textsuperscript{#1\if@tempswa , #2\fi}}
\makeatother

\renewcommand{\Im}{\text{Im}}

\newcommand{\mat}[1]{\begin{pmatrix}#1\end{pmatrix}}
\newcommand{\Dx}{\partial_x}

\newcommand{\DLT}[1][]{\mathcal{D}_{L/T#1}}
\newcommand{\DT}[1][]{\mathcal{D}_{T#1}}
\newcommand{\DL}[1][]{\mathcal{D}_{L#1}}
\newcommand{\T}[1][]{\mathcal{T}_{#1}}
\newcommand{\Tti}[1][]{\tilde{\mathcal{T}}_{#1}}
\newcommand{\jE}[1][]{j_{E#1}}
\newcommand{\jS}[1][]{j_{S#1}}
\newcommand{\jLT}[1][]{j_{L/T#1}}
\newcommand{\jL}[1][]{j_{L#1}}
\newcommand{\jT}[1][]{j_{T#1}}
\newcommand{\fLT}[1][]{f_{L/T#1}}
\newcommand{\fL}[1][]{f_{L#1}}
\newcommand{\fT}[1][]{f_{T#1}}
\newcommand{\ET}[1][]{E_{T#1}}

\newcommand{\const}[1]{\textrm{const}(#1)}
\newcommand{\integral}[1][]{\def\ArgI{{#1}}\IntegralRelay}
\newcommand\IntegralRelay[3][]{ \int_{\ArgI}^{#1}\!#3\,\mathrm{d}#2}
\newcommand{\mum}{\mathrm{\mu{}m}}
\newcommand{\cmss}{\mathrm{\frac{cm^2}{s}}}
\newcommand{\muVK}{\mathrm{\frac{\mu{}V}{K}}}
\newcommand{\nVK}{\mathrm{\frac{nV}{K}}}

\newcommand{\ETSNS}{E_{T,SNS}}
\newcommand{\LSNS}{L_{SNS}}
\newcommand{\RSNS}{R_{SNS}}
\newcommand{\MT}[1][]{\bigl<\DT[#1]^{-1}\bigr>}
\newcommand{\ML}[1][]{\bigl<\DL[#1]^{-1}\bigr>}
\newcommand{\MLT}[1][]{\bigl<\DLT[#1]^{-1}\bigr>}
\newcommand{\vc}[1]{\mathbf{#1}}
\newcommand{\Tr}{\text{Tr}}
\newcommand{\tanhe}[1]{\fL[,#1]^0}
\newcommand{\TT}[1][]{\bigl<\T[#1]\bigr>}
\newcommand{\TTti}[1][]{\bigl<\Tti[#1]\bigr>}
\newcommand{\E}{E}
\newcommand{\Q}{Q}
\newcommand{\QNN}{\Q_{\mathrm{NN}}}
\newcommand{\QNS}[1][]{{\Q_{\mathrm{NS#1}}}}
\newcommand{\RT}{\tilde{R}}
\newcommand{\G}{\check{G}}
\newcommand{\GA}{\hat{G}^A}
\newcommand{\GR}{\hat{G}^R}
\newcommand{\GK}{\hat{G}^K}

\newcommand{\tx}{\tau_1}

\newcommand{\tnz}{\hat{\tau}_3}
\newcommand{\tnv}{\hat{1}}

\newcommand{\tkv}{\check{1}}

\newcommand{\Sgin}{\check{\Sigma}_\text{in}}
\newcommand{\ff}{\mathfrak{f}}

\DeclareMathOperator{\sech}{sech}

\newcommand{\tmpnote}[1]%
   {\begingroup\color[named]{Red}\slshape(FIXME: #1)\endgroup}

\title{Thermopower in Andreev Interferometers}
\author{Pauli Virtanen and Tero T. Heikkil\"a}
\address{Low Temperature Laboratory, Helsinki University of
  Technology, \\ P.O. Box 2200 FIN-02015 HUT, Finland.}

\makeatletter
\runninghead{P. Virtanen and T. T. Heikkil\"a}{\@title}
\makeatother

\begin{document}
\maketitle

\begin{abstract}
 We examine the thermopower $\Q$ of a mesoscopic normal-metal (N) wire
in contact to superconducting (S) segments and show that even with
electron-hole symmetry, $\Q$ may become finite due to the presence of
supercurrents. Moreover, we show how the dominant part of $\Q$ can be
directly related to the equilibrium supercurrents in the structure. We
also discuss the thermopower arising due to an anomalous kinetic
coefficient which is finite in the presence of supercurrent and in
some situations gives the dominant contribution. In general, a finite
thermopower appears both between the N reservoirs and the
superconductors, and between the N reservoirs themselves. The latter,
however, strongly depends on the geometrical symmetry of the
structure. The paper includes a detailed analytical derivation of the
results and an exact numerical solution of the quasiclassical
equations in a few sample geometries.

\end{abstract}


\section{INTRODUCTION}

From time to time in the last century, the connection between the
thermoelectric effects and superconductivity has attracted
interest.\cite{ginzburgnobellecture} As early as 1927, Meissner
found that the thermopower is absent in a steady-state
superconductor. This is due to the fact that the thermoelectric
current is in this case counterbalanced by the supercurrent.
However, in 1944, Ginzburg \cite{ginzburg44} pointed out that
there can be other types of thermoelectric effects in
superconductors. Interestingly, many of them are still unobserved
\cite{ginzburgnobellecture}, at least partially due to the fact
that they are based on the same electron-hole asymmetry that
produces the thermoelectric effects in nonsuperconducting systems
\cite{ashcroft}, where Mott's law connects the thermopower to the
tiny difference of the effective masses between electrons and
holes (i.e., to the nonlinearity of the quasiparticle dispersion
relation above and below the Fermi sea). Such effects are governed
by the factor $k_B T/E_F$, and are thus very small in the low
temperatures required for superconductivity. Even after six
decades, the understanding of the experiments on some of these
effects is still lacking.\cite{vanharlingen82,galperin02} For
example, the measured thermoelectric flux in a bimetallic
superconducting ring was orders of magnitude larger than predicted
by the conventional theories.

Another effect of thermoelectric type, finite even in the presence
of complete electron-hole symmetry, was observed and successfully
explained in the turn of
1980's:\cite{pethick79,clarke79,schmid79,clarke80} it was found
that the combination of a temperature gradient and supercurrent
could make rise to a charge (branch) imbalance, i.e., a difference
in the electrochemical potentials of the normal and
superconducting components. The origin of this effect is similar
to that considered in this paper, but it manifests itself in quite
a different form in the systems studied here.

In the 1990's, many groups started to investigate the
superconducting effects induced in normal-metal (N) wires from
nearby superconductors (for review, see
Refs.~\onlinecite{raimondilambert,courtoispannetier,belzig99}).
For example, if the N metal is connected to two superconducting
segments, one can drive a supercurrent through it \cite{dubos01}.
It was found that this superconducting proximity effect influences
both the electrical \cite{charlat96,stoof96} and thermal
\cite{bezuglyi03} conductances of the N wire, in general breaking
the Wiedemann-Franz relation between the two.\cite{claughton96} In
most cases, these effects account for some tens of percent changes
in the conductances.

In proportion, a much greater effect was experimentally observed in
the thermopower \cite{eom98,dikin02,dikin02b,parsons03,parsons03b} of
"Andreev interferometers", structures where the normal-metal wire was
connected to two ends of a broken superconducting loop. The observed
thermopower was found to be orders of magnitude larger with the
proximity effect (below the superconducting critical temperature
$T_C$) than without it ($T>T_C$). A thermally induced voltage was
found both between the ends of the normal metal
\cite{eom98,dikin02,dikin02b,parsons03b} ("N-N thermopower" $Q_{NN}$)
as also between one end of the normal metal and the superconducting
contacts ("N-S thermopower" $Q_{NS}$).\cite{parsons03,parsons03b}
Moreover, it was found that this induced thermopower oscillates with
the flux applied through the superconducting loop.

It was shown in Ref.~\onlinecite{heikkila00}, including one of the
present authors, that Mott's law is not in general valid in the
presence of the proximity effect, indicating that the reason
underlying the observations is not necessarily dependent on
electron-hole symmetry. Indeed, the experimental observations were
partially explained, without invoking electron-hole asymmetry, in the
linear regime for $Q_{NS}$ by Seviour and Volkov \cite{seviour00} and
another effect was suggested for the presence of $Q_{NN}$ by Kogan,
Pavlovskii and Volkov.\cite{kogan02} The full nonlinear regime,
including important effects coming from asymmetries in the structure,
and a phenomenological explanation that connects the thermopower to
the temperature dependence of the supercurrent, was described by the
present authors in Ref.~\onlinecite{virtanen04}. This theory seems to be in
quantitative and qualitative agreement with most of the
observations. The aim of the present paper is to explain this theory
in detail and describe a related effect, thermopower due to an
anomalous kinetic coefficient, which is the main source of thermopower
in certain geometries.

Further, Ref.~\onlinecite{heikkila03} shows that the presence of
supercurrent can lead also to a Peltier-type effect. This effect
has not yet been experimentally observed.

This paper is organized as follows. In Sect.~\ref{sec:formalism}, we
detail the quasiclassical formalism applied in the calculation of the
thermopower. This formalism assumes complete electron-hole symmetry,
and the predicted effects are hence not limited by the factor $k_B
T/E_F$. The analytic approximations necessary for obtaining an
expression for the thermopower are presented in
Sect.~\ref{sec:thermopower}, including symmetry arguments for the full
(without approximations) solutions, a phenomenological picture of the
effect, and a calculation of the thermopower in the quasiequilibrium
limit. The general behavior of this thermopower is detailed in
Sect.~\ref{sec:general}, and effects related to the geometry of the
studied structures are discussed in Sect.~\ref{sec:geometry} Most of
the results in these sections are obtained numerically without any
approximations (beyond the quasiclassical diffusive limit).  Section
\ref{sec:comparison} includes a comparison of the theory to the
experiments and to the other theories on the effect. The main spectral
coefficients responsible for the thermopower, the spectral
supercurrent and the anomalous coefficient $\T$, are described in
Sect.~\ref{sec:spectralcoefs} Finally, the results are shortly
discussed in Sect.~\ref{sec:discussion} An impatient reader should
capture mostly the kinetic equations \eqref{eq:kinetic} in
Sect.~\ref{sec:formalism}, skip the detailed analytic derivations in
Subs.~\ref{sub:analytic} and proceed with the main results in
Subs.~\ref{sub:longjunction} and their characteristics in
Secs.~\ref{sec:general},\ref{sec:geometry}

\section{FORMALISM} \label{sec:formalism}


Many well-understood phenomena in inhomogeneous superconductivity have
been explained through the use of quasiclassical Green's
functions. This is in particular the case in mesoscopic
superconductivity (see the reviews in
Refs.~\onlinecite{raimondilambert,belzig99}). The essence of the
quasiclassical theory is ignoring the short-scale oscillations in the
relative coordinate, and concentrating only on the center-of-mass
coordinate of the space-dependent Green's functions. Such an
approximation ignores the quantum-mechanical interference effects
arising from short scales of the order of the Fermi wavelength
(structural variations on this scale can be included through boundary
conditions, see below and Ref.~\onlinecite{nazarov99}), and the
effects related to the nonlinear parts of the quasiparticle dispersion
relation (the theory is based on the linearisation of the dispersion
relation around the Fermi energy). Therefore, it cannot describe the
normal-metal thermoelectric effects which arise from the energy
dependence of the effective mass.

In this paper, we concentrate on the diffusive limit, where all the
important length scales are much greater than the mean free path. This
is typically the case in metallic wires. To describe nonequilibrium
effects (finite voltage) in the nonlinear regime, we adopt the Keldysh
formalism.\cite{rammer86} In the considered limit, the
Keldysh--Green's function is described by the Usadel
equation.\cite{usadel70}

\subsection{Keldysh--Usadel equations}

The Keldysh--Usadel equation for the Green's functions
$\G(\vec{r},\E)$ in the Keldysh~$\otimes$~Nambu space can be written
in the static case as
\begin{equation}\label{eq:ku}
  \nabla\cdot(D \G \nabla \G) = \left[-i \E \; 1\otimes\tnz + \check\Delta
    + \Sgin, \G \right] \,.
\end{equation}
Here, $D$ is the diffusion constant of the metal, $\Sgin$ contains
the inelastic scattering self-energies, and $\check\Delta$ is the
pair potential matrix. The pair potential is self-consistently
related to the solutions of Eq.~\eqref{eq:ku}, \cite{belzig99} but
it vanishes in normal metals, on which we concentrate here
(superconductivity is induced by a boundary condition from bulk
superconductors). Moreover, $\G$ has the matrix structure
\begin{gather}
  \G = \mat{ \GR & \GK \\ 0 & \GA }
  \,,\;
  \GA = - \tnz (\GR)^\dagger \tnz
  \,,\;
  \GK = \GR \hat{h} - \hat{h} \GA
  \,,
\end{gather}
where $\GR$, $\GA$ and $\GK$, the retarded, advanced and Keldysh
components, are $2\times2$ matrices, and the last relation is due to
the normalization $\G^2 = \tkv$. The electron--hole distribution
function matrix $\hat{h}$ introduced here has two free parameters and
it can be chosen of the form $\hat{h}\equiv\fL\tnv + \fT\tnz$, where
$\fT$ and $\fL$ are the symmetric and antisymmetric parts,
\begin{gather}\label{eq:fTfL}
  \fT(\E) \equiv 1 - \ff(\mu_S - \E) - \ff(\mu_S + \E), \;
  \fL(\E) \equiv \ff(\mu_S - \E) - \ff(\mu_S + \E)
  \,
\end{gather}
of the electron distribution function $\ff(\vec{r},\E)$ and the
energy $\E$ is given with respect to the potential $\mu_S$ of the
superconductors.

The Green's function obtained from Eq.~\eqref{eq:ku} may be used for
evaluating the observable energy and charge current densities $j_Q$
and $j_c$:
\begin{subequations}
\begin{align}
  j_Q &= \frac{\sigma}{2e^2} \integral{\E}{\E\jL} \,,
  &
  j_c &= \frac{\sigma}{2e} \integral{\E}{\jT} \,,
  \\
  \jL &\equiv \frac{1}{4} \Tr[(\tx\otimes\tnv)\,\G\nabla\G\,] \,,
  &
  \jT &\equiv \frac{1}{4} \Tr[(\tx\otimes\tnz)\,\G\nabla\G\,] \,.
\end{align}
\end{subequations}
Here, $\jL$ and $\jT$ are the spectral current densities, and $\sigma$
is the normal-state conductivity of the metal.

It is useful to parameterize $\G$ so that it automatically
satisfies the normalization. Since $\G^2=\tkv$ implies that
$(\GR)^2=\tnv$, one can parametrize $\GR$ in terms of two complex
functions $\theta$ and $\chi$ as
\begin{equation}
  \GR = \mat{
    \cosh(\theta)               &    e^{i \chi} \sinh(\theta) \\
    -e^{-i \chi} \sinh(\theta)   &  -\cosh(\theta)
  }
  \,,
\end{equation}
and hence $\GA$ and $\GK$ depend only on $\theta$, $\chi$, $\fT$ and
$\fL$.  To find the equations that determine $\theta$ and $\chi$, one
can consider the retarded part in Eq.~\eqref{eq:ku}, and, neglecting
inelastic effects ($\Sgin=0$), obtain the spectral Keldysh--Usadel
equations in normal metals ($\check\Delta=0$)
\begin{subequations}
  \label{eq:spectral}
  \begin{gather}
    D \nabla^2 \theta = -2 i \E \sinh(\theta)
    + \frac{1}{2} D (\nabla \chi)^2 \sinh(2\theta) \,,
    \label{eq:spectral1}
    \\
    D \nabla\cdot \jE = 0,
    \quad \jE \equiv -\sinh^2(\theta) \nabla\chi \,.
    \label{eq:spectral2}
  \end{gather}
\end{subequations}
These equations describe the behavior of the superconducting proximity
effect: the pairing amplitude can be written as
$F=\sinh(\theta)\exp(i\chi)$, which shows that $\theta$ is related to
the strength of the proximity effect and $\chi$ to the phase of
the superconducting order parameter.

Using the same assumptions as above and taking the traces
$\Tr[(\tx\otimes\hat{1})\;\cdot\;]$ and
$\Tr[(\tx\otimes\tnz)\;\cdot\;]$ of Eq.~\eqref{eq:ku}, one obtains the
kinetic equations for the spectral energy and charge current densities
\begin{subequations}
  \label{eq:kinetic}
  \begin{align}
    D\nabla\cdot \jL &= 0, &
    \jL = \DL\nabla\fL - \T\nabla\fT + \jS\fT
    \,,
    \label{eq:kinetic1}
    \\
    D\nabla\cdot \jT &= 0, &
    \jT = \DT\nabla\fT + \T\nabla\fL + \jS\fL
    \,.
    \label{eq:kinetic2}
  \end{align}
\end{subequations}
These describe the kinetics of electrons and holes in the presence of
the proximity effect, and, moreover, imply that $\jL$ and $\jT$ are
conserved. The spectral coefficients in Eqs.~\eqref{eq:kinetic} are
determined from the solutions to Eqs.~\eqref{eq:spectral}:
\begin{subequations}
  \label{eq:coefficients}
  \begin{align}
    \DL &\equiv \frac{1}{4} \Tr[1\!-\!\GR\GA]
    = \frac{1}{2}(1+|\cosh\theta|^2-|\sinh\theta|^2\cosh(2\Im[\chi])),
    \label{eq:DL}
    \\
    \DT &\equiv \frac{1}{4} \Tr[1\!-\!\GR\tnz\GA\tnz]
    = \frac{1}{2}(1+|\cosh\theta|^2+|\sinh\theta|^2\cosh(2\Im[\chi])),
    \label{eq:DT}
    \\
    \T  &\equiv \frac{1}{4} \Tr[\GA\GR\tnz]
    = \frac{1}{2}|\sinh\theta|^2\sinh(2\Im[\chi])),
    \\
    \jS &\equiv \frac{1}{4} \Tr[(\GR\nabla\GR - \GA\nabla\GA)\tnz]
    = \Im[-\sinh^2(\theta)\nabla\chi] = \Im[\jE].
    \label{eq:T}
  \end{align}
\end{subequations}
Here, $\DL$ and $\DT$ are the spectral energy and charge diffusion
coefficients, and $\jS$ is the spectral density of the
supercurrent-carrying states. \cite{heikkila02scdos} The
cross-term $\T$ is usually small but not completely negligible.

To simplify the problem, one often assumes that wires are quasi-1D
structures, i.e., much longer than wide, and translationally invariant
in their latitudinal directions. Thus, one needs only to consider
longitudinal variation in the quantities, and the gradients may be
replaced with 1D-derivatives.

\subsection{Boundary conditions}

The Keldysh--Usadel equations cannot handle changes in the structure
which occur in a distance short compared to the Fermi wavelength or
the mean free path, so boundary conditions describing interfaces and
junctions need to be derived using different techniques.

To obtain the boundary conditions at nodes of one-dimensional wires
with clean metallic contacts, one may use the the matrix circuit
theory. \cite{nazarov99} It yields a Kirchhoff-like conservation law
for the spectral matrix currents $A\sigma\G\Dx\G$, where $A$ is the
cross-sectional area of a wire. In the $\theta$-parameterized form,
this results in
\begin{subequations}
  \label{eq:current-conservation}
  \begin{align}
    \sum_i A_i \sigma_{i} \nabla \theta_i &= 0, &
    \sum_i A_i \sigma_{i} \nabla \chi_i &= 0,
    \label{eq:current-conservation1}
    \\
    \sum_i A_i \sigma_{i} \jL &= 0, &
    \sum_i A_i \sigma_{i} \jT &= 0,
  \label{eq:current-conservation2}
  \end{align}
\end{subequations}
and to the continuity of $\theta$, $\chi$, $\fL$ and $\fT$. In
Eqs.~\eqref{eq:current-conservation}, the sums go over the wires
ending in the node.

The properties of connections between wires and terminals are also
described by a matrix conservation condition. \cite{nazarov99} In the
limit of a clean metallic contact, at the interface to a normal
reservoir, Green's function $\G$ and thus also $\theta$, $\chi$, $\fL$
and $\fT$ get their bulk values:
\begin{subequations}
  \label{eq:Nboundary}
  \begin{align}
    \theta &= 0, \quad \chi = \textrm{arbitrary}, \\
    \fLT &= \frac{1}{2} \left( \tanh\left(\frac{\E+\mu}{2 k_B T}\right) \pm
                             \tanh\left(\frac{\E-\mu}{2 k_B T}\right)\right),
  \end{align}
\end{subequations}
where $T$ is the temperature of the reservoir and $\mu$ its potential,
compared to $\mu_S$. Thus, near normal metal reservoirs or far from
superconductors, the coefficients~\eqref{eq:coefficients} obtain their
normal-state bulk values $\DL\rightarrow 1$, $\DT\rightarrow 1$ and
$\T\rightarrow 0$.

At a clean metallic contact to a superconducting reservoir, bulk
values are again obtained, with the exception of $\fL$ for energies
inside the superconducting energy gap $\Delta$:
\begin{subequations}
  \label{eq:Sboundary}
  \begin{align}
    &\theta = \textrm{artanh}\frac{\Delta}{\E}, \quad \chi = \phi,
    \quad \fT = 0, \\
    |\E| > \Delta:\quad& \fL = \tanh\left(\frac{\E}{2 k_B T}\right), \\
    |\E| < \Delta:\quad& \jL = 0
    \,.
  \end{align}
\end{subequations}
Here, $\phi$ is the phase of the superconducting order parameter.  We
assumed here no charge imbalance, $\mu_S=0$, hence $\fT=0$.  Moreover,
the boundary conditions for $\theta$ and $\chi$ imply that near a
superconductor interface $\DL\rightarrow 0$ and $\T\rightarrow 0$ for
energies $|\E|<\Delta$, which in fact implies that $\jL = 0$. Thus no
contribution to the energy current density $j_Q$ through the interface
comes from inside the energy gap --- this is a known property of the
Andreev reflection.

With these boundary conditions, the spectral
equations~\eqref{eq:spectral} depend only on the phase differences and
the geometry of the system, while kinetic equations~\eqref{eq:kinetic}
depend also on the temperatures and the potentials of the
reservoirs. The system is, however, fully decoupled from the
temperatures of the superconductors for the energies $|\E| < \Delta$.

Note that in wires connected to a normal reservoir, the spectral
supercurrent $\jE\equiv 0$, due to the boundary condition
$\theta=0$. Since generally $\sinh(\theta) \not\equiv 0$ in these
wires, the conservation of spectral supercurrent $\jS$ implies the
conservation of the phase, thus $\chi=\const{x}$. Similarly, in wires
connected to a superconducting reservoir, there is the boundary
condition $\jL \equiv 0$ for $|\E|<\Delta$, and one can eliminate
$\fL$ from the kinetic equations.

\subsection{Energy and length scales}

Scaling the spatial dimensions in the spectral equations
\eqref{eq:spectral} by the length $L$ (where $L$ can be one of the
wire lengths, or most naturally for the supercurrent, the distance
between the superconductors), one finds a natural energy scale: the
Thouless energy
\begin{equation}\label{eq:E_T}
  \ET \equiv \frac{\hbar D}{L^2}
  \approx \left\{
  \begin{array}{cc}
    13~\mu \text{V} & \; e \\
    0.15~\text{K}   & \; k_B
  \end{array}
  \right\} \;
  \frac{D/(200~\cmss)}{(L/\mum)^2}
  \,,
\end{equation}
This energy is the inverse time of flight through a diffusive wire of
length $L$.

In practice, there are two different types of energy scales
describing the behavior of the spectral functions: $E_T$ and the
superconducting energy gap $\Delta$ of the electrodes connected to
the normal-metal wire. \cite{heikkila02scdos} The previous is
important in the case of long junctions $L \gg \xi_0$ (i.e., $E_T
\ll \Delta$) and the latter for $L \lesssim \xi_0$ ($E_T \gtrsim
\Delta$). Here $\xi_0=\sqrt{\hbar D/2\Delta}$ is the
superconducting coherence length. For example, for $T \ll T_C$, in
Al $\xi_0 \approx 200$~nm, and in Nb $\xi_0 \approx 30$~nm.

In the long-junction limit $\Delta\gg\ET$, the energy scale for
the solution to~(\ref{eq:spectral},\ref{eq:kinetic}) with the
boundary conditions
(\ref{eq:current-conservation}-\ref{eq:Sboundary}) depends only on
$E_T$. Thus, by changing $\ET$, the results may be scaled to fit
all systems with similar ratios of wire lengths, cross sectional
areas and conductivities.

\section{THERMOPOWER IN ANDREEV INTERFEROMETERS} \label{sec:thermopower}

A temperature difference $\Delta T$ between two points in a metal
may induce a charge current $I_c$. Disconnecting the system from
its surroundings makes the current vanish, and a potential
difference $\Delta V$ arises instead. The thermopower $Q$ is the
ratio of this potential difference to the temperature difference,
$Q \equiv \left(\Delta\!V / \Delta\!T\right)_{I_c=0}$. In this
section, we show how the supercurrent flowing in Andreev
interferometers connects the temperatures and potentials, inducing
a finite thermopower.

\subsection{Thermopower without superconductivity}

An estimate for the thermopower in the absence of
superconductivity can be found using the semiclassical Boltzmann
equation as, for example, in Ref.~\onlinecite{ashcroft}. Such a
description can be obtained from the microscopic theory presented
above, provided that the length scales important for the problem
are much greater than the Fermi wavelength.\cite{rammer86} This
analysis leads to the Mott relation for the thermopower
\begin{equation}\label{eq:mott}
  Q = -\frac{\pi^2}{3} \, \frac{k_B^2 T}{e} \,
  \left.
  \frac{\mathrm{d}\ln{\sigma(\E)}}{\mathrm{d}\E}
  \right|_{\E=\E_F}
  \,.
\end{equation}
Here, $\sigma(\E)$ describes a generalized conductivity ---
$\sigma(\E_F)$ gives the actual conductivity in the linear-response
regime --- and the logarithmic derivative depends on the energy
dependence of the elastic relaxation time and the exact structure of
the Fermi surface. The previous can typically be neglected, and the
latter is due to the small asymmetry between the dispersion relations
for electrons and holes. Both of these effects are neglected in the
quasiclassical theory. There are also other major contributions to the
thermopower, e.g. the phonon drag, but for simple metals, they should
vanish at high and very low temperatures. \cite{hansch84}

We can use the temperature dependence in Eq.~\eqref{eq:mott} for an
extrapolation from the room-temperature values of $Q$ to its value at
subkelvin temperatures. Usual magnitudes for the thermopower at the
room temperature are of the order $\muVK$ for pure metals. To obtain a
rough estimate at a low temperature, we extrapolate $10~\muVK$ from
room temperature to a temperature of 0.5~K (typical for the studies on
the proximity effect), which yields $Q \sim 10~\nVK$. Measurements in
Ref.~\onlinecite{parsons03} gave a stricter estimate $Q\lesssim1~\nVK$
for their case.

However, the Keldysh--Usadel equations predict the presence of a
large thermopower $Q\sim\muVK$, even under complete electron-hole
symmetry, as supercurrents couple the temperatures to the
potentials (through the coefficients $\jS$ and $\T$ in the kinetic
equations~\eqref{eq:kinetic}).

\subsection{Modeled structure}

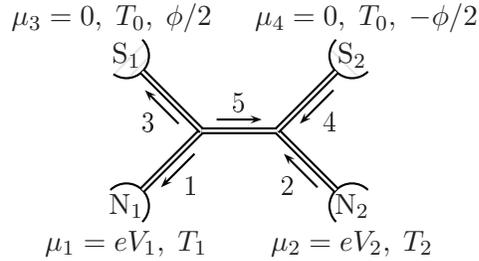
\begin{figure}\centering
  \begin{pspicture}(-1.5, -.6)(4.5, 2.6)
  \psset{xunit=1cm,yunit=1cm}
  \SpecialCoor
  
  \newcommand{\mrad}{.3}
  \rput(0,0){\cnode[linestyle=none]{\mrad}{N1}}
  \rput(3,0){\cnode[linestyle=none]{\mrad}{N2}}
  \rput(0,2){\cnode[linestyle=none]{\mrad}{S1}}
  \rput(3,2){\cnode[linestyle=none]{\mrad}{S2}}
  \rput(1,1){\rnode{C1}{\pscircle*{0.05}}}
  \rput(2,1){\rnode{C2}{\pscircle*{0.05}}}
  
  \rput(N1){$\mathrm{N}_1$} \psarc(N1){\mrad}{-45}{135}
  \rput(N2){$\mathrm{N}_2$} \psarc(N2){\mrad}{ 45}{225}
  
  \newgray{ltgr}{.90}
  \psarc[fillstyle=hlines,hatchcolor=ltgr](S1){\mrad}{-125}{45}
  \rput(S1){$\mathrm{S}_1$} 
  \psarc[fillstyle=vlines,hatchcolor=ltgr](S2){\mrad}{-225}{-45}
  \rput(S2){$\mathrm{S}_2$} 

  \rput(0,-.5){$\mu_1=eV_1,\; T_1$}
  \rput(3,-.5){$\mu_2=eV_2,\; T_2$}
  \rput(0,2.5){$\mu_3=0,\; T_0,\;\phi/2\quad$}
  \rput(3,2.5){$\quad\mu_4=0,\; T_0,\;-\phi/2$}

  \psset{labelsep=.08,offset=.15,nodesep=.2}
  \ncline{->}{C1}{N1} \Aput{1}
  \ncline{->}{N2}{C2} \Aput{2}
  \ncline{->}{C1}{S1} \Aput{3}
  \ncline{->}{S2}{C2} \Aput{4}
  \ncline{->}{C1}{C2} \Aput{5}
  \psset{offset=0,nodesep=0}
  
  \psset{doubleline=true}
  \ncline{-}{N1}{C1} 
  \ncline{-}{C2}{N2} 
  \ncline{-}{S1}{C1} 
  \ncline{-}{C2}{S2} 
  \ncline{-}{C1}{C2} 
\end{pspicture}
  \caption{The system under examination. The terminals S are
    superconducting reservoirs, N normal-metal reservoirs and they are
    connected by normal-metal wires with lengths $L_k$, cross sections
    $A_k$ and normal-state conductivities $\sigma_k$, $ k=1 \ldots
    5$. Temperatures $T$ and potentials $\mu=eV$ in the reservoirs are
    as shown. Moreover, all contacts are assumed clean
    metallic. Positive directions for currents are indicated with
    arrows.}
  \label{fig:kinetic-wires}
\end{figure}

Here we focus to study thermopower in the structure shown in
Fig.~\ref{fig:kinetic-wires}, where two superconducting (S) reservoirs
with a phase difference $\phi$ are connected through a normal-metal
link to two normal metal (N) reservoirs. When measuring thermopower in
this structure, no current flows in wires 1 and 2,
\begin{equation}\label{eq:nc}
  j_{c,1} = 0,\quad j_{c,2} = 0,
\end{equation}
and definite voltages $V_1$ and $V_2$ are induced in the N
reservoirs. Here we may define several thermopower-like quantities:
\begin{equation}
  \QNN     = \frac{V_2 - V_1}{T_2 - T_1}\,,\quad
  \QNS[,1] = \frac{V_1}{T_2 - T_1}\,,\quad
  \QNS[,2] = \frac{V_2}{T_2 - T_1}\,.
\end{equation}
Clearly, their behavior with varying parameters need not be
similar. That is, generally the potential induced in the colder
electrode may differ from that induced to the hotter electrode.

The thermopower appearing in the structure may be estimated by
applying the formalism described in Sect.~\ref{sec:formalism} We have
solved Eqs.~(\ref{eq:spectral},\ref{eq:kinetic}) numerically, and
derived analytic approximations for the solutions of the kinetic
equations~\eqref{eq:kinetic}. In the following, we first discuss the
analytic results, as they provide some insight to the problem, and
then examine interesting features numerically (Secs.~\ref{sec:general}
and \ref{sec:geometry}).

\subsection{Qualitative picture}

The effect of supercurrents on the thermopower can be understood
phenomenologically as follows. If $T_1 \ne T_2$, the
temperature-dependent \cite{dubos01} equilibrium supercurrent
$I_S(T_1)$ in wire 3 is different from $I_S(T_2)$ in wire 4. (For this
qualitative picture, we approximate these wires to be at the
temperatures $T_1$, $T_2$ of the normal reservoirs.) Conservation of
currents should still be maintained, so a compensating effect must
arise. Should the normal reservoirs be kept at the same potential as
the superconductors, a quasiparticle current $I_{\text{qp}} \propto
(I_S(T_1)-I_S(T_2))$ from them to the superconductors would balance
the difference.  However, when no current is allowed to flow in wires
1 and 2, a compensating N--S potential difference $V_N-V_S \propto R
(I_S(T_1) - I_S(T_2))$ is induced instead.

The qualitative picture gives the following predictions: First, the
induced potentials should oscillate with the phase difference $\phi$
between the S electrodes, similarly to the supercurrent. Moreover, the
potentials induced in the N electrodes need not be the same,
especially if the resistances of the wires are not symmetric at the
left and at the right. Thus, the supercurrent-effect also contributes,
as in Eq.~\eqref{eq:nnapprox}, to the thermopower $\QNN$ between the N
electrodes. Such a contribution should be present even in structurally
symmetric setups, as the resistances are temperature-dependent due to
the proximity effect. \cite{charlat96}

\subsection{Phase behavior} \label{sub:phasedep}

If the phase difference between the superconducting elements vanishes,
the quasiclassical equations predict a vanishing thermopower $Q$ (as
shown below). In the presence of a phase difference, the supercurrent
and the term $\T$ couple the two types of distribution functions,
inducing a finite $Q$. Although the solution to Eq.~\eqref{eq:ku}
cannot be found in closed form, we can obtain information about the
exact phase behavior of the thermopower using symmetry
considerations. Suppose we have the solutions $\chi$, $\theta$, $\fL$
and $\fT$ for the Keldysh--Usadel
equations~(\ref{eq:spectral},\ref{eq:kinetic}) in the system. Then,
define
\begin{equation}
  \tilde\chi   \equiv -\chi     \,,\quad
  \tilde\theta \equiv \theta    \,,\quad
  \tilde\fL    \equiv \fL   \,,\quad
  \tilde \fT   \equiv -\fT
  \,.
\end{equation}
First, we see that these new variables are solutions to
Eqs.~\eqref{eq:spectral}. This implies that $\tilde\DT=\DT$,
$\tilde\DL=\DL$, $\tilde\T=-\T$ and $\tilde\jS=-\jS$, so the new
variables are also solutions to Eqs.~\eqref{eq:kinetic}. They also
satisfy the spectral current
conservation~\eqref{eq:current-conservation}, with $\tilde\jT=-\jT$
and $\tilde\jL=\jL$, and the boundary
conditions~(\ref{eq:Nboundary},\ref{eq:Sboundary}), but with
$\tilde\phi=-\phi$ and $\tilde\mu = -\mu$. Thus, there is a solution
with inverted charge currents $I_c$, potentials $\mu$ and phases
$\phi$. For the thermopower this implies that
\begin{equation}
  Q(\phi) = \left.\frac{\Delta V}{\Delta T}\right|_{I_c=0}
          = -\left.\frac{\Delta \tilde V}{\Delta \tilde T}\right|_{\tilde I_c=0}
      = -Q(-\phi)
  \,.
\end{equation}
Moreover, $Q$ is necessarily $2\pi$-periodic in phase, so this implies
that it should also be antisymmetric with respect to $\phi=n\pi$ for
all $n\in\mathbb{Z}$.

The result also implies that in a system with no phase differences or
no superconducting parts, the thermopower vanishes. This is in
agreement with the Mott-law prediction due to the quasiclassical
approximation. However, note that the reasoning above assumes that the
condition~\eqref{eq:nc} determines the solution uniquely and that the
thermopower is well defined. This may not be the case in some
pathological geometries in the presence of the proximity effect, as
the condition~\eqref{eq:nc} is determined only for the
energy-integrated currents. In the cases studied in the present
manuscript, the convergence of numerics hints that the solutions are
unique.

\subsection{Analytic approximations} \label{sub:analytic}

Although we cannot solve analytically the spectral equations in the
structure of Fig.~\ref{fig:kinetic-wires}, we can obtain useful
approximations starting from the kinetic equations.

As a first step, we assume that the induced potentials are small
compared to the temperature, $|\mu| \ll k_B T$. Since $\fT$ is related
to the potentials,
we can reason that $\fT$ is small, and hence neglect the terms
$-\T\partial_x\fT$ and $\jS\fT$ in Eq.~\eqref{eq:kinetic1}. Comparison
with numerics shows that this is a good approximation.

The kinetic equations, where inelastic effects are neglected, can now
be integrated over a wire of length $L$. This relates the currents to
the distribution functions $\fL$, $\fT$ at the ends of the wire:
\begin{subequations}
  \label{eq:approxjLT}
  \begin{align}
    \jL &= \frac{\fL(L) - \fL(0)}{L \ML} \label{eq:approxjL}
    \,,
    \\
    \jT &= \label{eq:approxjT}
    \frac{\fT(L) - \fT(0)}{L \MT}
    + \jS \fL(0)
    \\\notag
    &+ \frac{\fL(L) - \fL(0)}{
      L^2 \ML \MT} \!\left( \integral[0][L]{x}{\!\!\!\frac{\T}{\DL \DT}}
    + \jS \int_0^L \!\!\! \int_0^x \!\!\! \frac{\mathrm{d}x' \mathrm{d}x}{\DL(x') \DT(x)}
    \right)
    \,.
  \end{align}
\end{subequations}
Here $\MLT \equiv \frac{1}{L} \int_0^L \DLT^{-1}\,\mathrm{d}x$ are the
dimensionless spectral energy and charge resistances. All the
coefficients can be calculated numerically by solving the spectral
equations. We see that Eq.~\eqref{eq:approxjLT} is linear in the
distribution functions, so we can use the continuity of $\fT$ and
$\fL$ at the nodes and write the Kirchhoff-like spectral current
conservation equations~\eqref{eq:current-conservation2} for the
structure in Fig.~\ref{fig:kinetic-wires} as:
\begin{subequations}
  \label{eq:conservation-in-struct}
  \begin{align}
    \sigma_1 A_1 \vc{X_1} \mat{ \vc{f}_5(0) \\ \vc{f}_1^0 }
    + \sigma_3 A_3 \vc{X_3} \mat{ \vc{f}_5(0) \\ \vc{f}_1^0 }
    + \sigma_5 A_5 \vc{X_5} \mat{ \vc{f}_5(0) \\ \vc{f}_5(L) } &= \vc{0}
    \,,
    \\
    \sigma_2 A_2 \vc{X_2} \mat{ \vc{f}_2^0  \\ \vc{f}_5(L) }
    + \sigma_4 A_4 \vc{X_4} \mat{ \vc{f}_4^0  \\ \vc{f}_5(L) }
    + \sigma_5 A_5 \vc{X_5} \mat{ \vc{f}_5(0) \\ \vc{f}_5(L) } &= \vc{0}
    \,.
  \end{align}
\end{subequations}
Here, $\vc{f} \equiv (\fL,\fT)$, and $\vc{f}_k^0$ contain the electron
distribution function parts in the reservoir $k$, as in
Eqs.~(\ref{eq:Nboundary}, \ref{eq:Sboundary}). Moreover, the matrices
$\vc{X}_k$ can be obtained using Eqs.~\eqref{eq:approxjLT} --- noting
that the results for $|\E|<\Delta$ and $|\E|>\Delta$ are different due
to the differing boundary conditions in the kinetic equations.
Equation~\eqref{eq:conservation-in-struct} is linear, and the unknown
$\vc{f}_5(0)$ and $\vc{f}_5(L)$ may be solved in a straightforward
manner. Thus, for given temperatures and potentials in the reservoirs,
we obtain the spectral currents $\jL$ and $\jT$ for each wire.

To determine the thermopower, we have to find $\mu_1$ and $\mu_2$ such
that the condition~\eqref{eq:nc} is satisfied. These potentials are
expected to be small, so we can linearize the distribution functions
in the reservoirs with respect to them,
\newcommand{\linfT}[1]{\sech^2\left(\frac{\E}{2 k_B T_{#1}}\right) %
  \frac{\mu_{#1}}{2 k_B T_{#1}}}
\newcommand{\linfTnomu}[1]{\sech^2\left(\frac{\E}{2 k_B T_{#1}}\right) %
  \frac{1}{2 k_B T_{#1}}}
\newcommand{\linfL}[1]{\tanh\left(\frac{\E}{2 k_B T_{#1}}\right)}
\begin{equation}
  \label{eq:linearizef}
  \vc{f}_k^0 \simeq \mat{ \tanhe{k} \\ \fT[,k]^0 }
    \equiv \mat{\linfL{k} \\ \linfT{k}} \,, \quad k=1,2\,,
\end{equation}
in the expressions for $j_{c,1}$ and $j_{c,2}$ obtained from
Eq.~\eqref{eq:conservation-in-struct}. Thus, we obtain a linear
equation for the potentials $\mu_1$ and $\mu_2$, which can then be
solved.

However, without further approximations, the final expression for the
thermopower is too complicated to provide much insight. We make the
following simplifications:
\begin{enumerate}
\item We approximate $\DL$ with its normal-state value $\DL=1$. This
  is a good approximation only away from superconductor interfaces,
  but numerics show that it usually has little influence on the
  thermopower.
\item Similarly, we approximate $\DT\approx1$ in the latter term in
  Eq.~\eqref{eq:approxjT}, and retain it only in the first term. This
  is because $\T$ and $\jS$ provide more essential energy-dependent
  features than $\DT$, but we still wish to retain the temperature
  dependence of conductance due to the coefficient in the first term.
  Comparing to exact numerics, we see that these approximations
  do not essentially affect the resulting thermopower.
\end{enumerate}

Using the above and the conservation of $\sigma A\jS$ we can write
Eqs.~\eqref{eq:conservation-in-struct} for the structure in
Fig.~\ref{fig:kinetic-wires} in a useful form at $|E| < \Delta$ ---
noting that $A\sigma\jS=A_3\sigma_3\jS[3]=-A_5\sigma_5\jS[5]$
according to the chosen directions:
\begin{subequations}
\begin{gather}
  \vc{M}^{-1} \equiv
  \mat{
    g_1 + g_5 & 0 & - g_5 & 0
    \\
    -\frac{A\sigma\jS}{2} + g_1 \TT[1] + g_5 \TT[5] &
    \tilde g_{135} &
    \frac{A\sigma\jS}{2} - g_5 \TT[5] &
    -\tilde g_5
    \\
    -g_5 & 0 & g_2 + g_5 & 0
    \\
    -\frac{A\sigma\jS}{2} - g_5 \TT[5] &
    -\tilde g_5 &
    \frac{A\sigma\jS}{2} + g_2 \TT[2] + g_5 \TT[5] &
    \tilde g_{245}
  }
  \,,
  \\
  \vc{M}^{-1}
  \underbrace{
    \mat{ \fL[,5](0) \\ \fT[,5](0) \\ \fL[,5](L) \\ \fT[,5](L) }
  }_{\vc{f}}
  =
  \underbrace{
  \mat{
    g_1 & 0 & 0 & 0 \\
    g_1 \TT[1] & \tilde g_1 & 0 & 0 \\
    0 & 0 & g_2 & 0 \\
    0 & 0 & g_2 \TT[2] & \tilde g_2
  }
  }_{\vc{C}}
  \underbrace{
    \mat{ \fL[,1]^0 \\ \fT[,1]^0 \\ \fL[,2]^0 \\ \fT[,2]^0 }
  }_{\vc{f}_{1,2}}
  \,,
\end{gather}
\end{subequations}
where $g_k = \sigma_k A_k / L_k$ is the conductance of wire $k$,
$g_{ijk} = g_i+g_j+g_k$, $\tilde g_k \equiv g_k / \MT[,k]$ and
$\TT[k] \equiv \frac{1}{L_k} \integral[0][L_k]{x}{\T[k]\,}$. For
$|E|>\Delta$, the result is
\begin{subequations}
\begin{gather}
  \vc{M}_\Delta^{-1} \equiv
  \mat{
    g_{135} & 0 & - g_5 & 0
    \\
    g_1\TT[1]\!+\!g_3\TT[3]\!+\!g_5\TT[5] &
    \tilde g_{135} &
    \frac{A\sigma\jS}{2} - g_5 \TT[5] &
    -\tilde g_5
    \\
    -g_5 & 0 & g_{245} & 0
    \\
    -\frac{A\sigma\jS}{2} - g_5 \TT[5] &
    -\tilde g_5 &
    g_2\TT[2]\!+\!g_4\TT[4]\!+\!g_5\TT[5] &
    \tilde g_{245}
  }
  \,,
  \\
  \vc{M}_\Delta^{-1} \vc{f}
  =
  \vc{C} \vc{f}_{1,2}
  +
  \underbrace{
    \mat{
      g_3 & 0 \\
      \frac{A\sigma\jS}{2}\!+\!g_3\TT[3] & 0 \\
      0 & g_4 \\
      0 & -\frac{A\sigma\jS}{2}\!+\!g_4\TT[4]
    }
  }_\vc{D}
  \underbrace{
    \mat{ \fL[,3]^0 \\ \fL[,4]^0 }
  }_{\vc{f}_{0}}
  \,,
\end{gather}
\end{subequations}
Note that the temperature of the superconductors enters for energies
$|E|>\Delta$ in the distribution functions $\fL[,3/4]^0$.

Using Eq.~\eqref{eq:approxjLT} with the assumption as above, we can
calculate the currents flowing in wires 1 and 2:
\begin{equation}
  \label{eq:analcurrents}
  \mat{
    -A_1\sigma_1 \jL[,1] \\ -A_1\sigma_1 \jT[,1] \\
     A_2\sigma_2 \jL[,2] \\  A_2\sigma_2 \jT[,2]
  }
  =
  \begin{cases}
    \vc{C} (\vc{M}\vc{C} - \vc{I}) \, \vc{f}_{1,2}\,,
                                     & \text{for $|E|<\Delta$} \\
    \vc{C} (\vc{M}_\Delta\vc{C} - \vc{I}) \, \vc{f}_{1,2}
    + \vc{C} \vc{M}_\Delta\vc{D} \, \vc{f}_{0} \,, & \text{for $|E|>\Delta$}
  \end{cases}
  \,.
\end{equation}
Here, the sign of $\jLT[,1]$ is due to the choice of directions in
Fig.~\ref{fig:kinetic-wires}. Now, we can integrate the rows 2 and 4
over the energy to obtain the observable charge currents, and
linearize the distribution functions. Condition~\eqref{eq:nc} then
yields an expression of the form
\begin{equation} \label{eq:analapproxpot}
  \integral[0][\infty]{\E}{ \vc{g}\, } \mat{\mu_1 \\ \mu_2} =
  \integral[0][\Delta]{\E}{
    \vc{P}_1 \mat{ \tanhe{1} \\ \tanhe{2} }
  }
  +
  \integral[\Delta][\infty]{\E}{
    \left[
    \vc{P}_2 \mat{ \tanhe{1} \\ \tanhe{2} }
    +
    \vc{P}_3 \mat{ \tanhe{0} \\ \tanhe{0} }
    \right]
  }
  \,.
\end{equation}
Here, $\vc{g}$ is a matrix containing temperature-dependent
conductance-like quantities, and the integrals with $\vc{P}_k$ yield,
for example, terms proportional to the supercurrents flowing in the
system at different temperatures. In the following, we simplify the
expression further to obtain the dominant features. However,
Eq.~\eqref{eq:analapproxpot} can well be used to obtain numerical
thermopower estimates for given spectral data, without further
approximations.

\subsection{Long-junction limit} \label{sub:longjunction}

In this section, we discuss the dominant features of the thermopower,
in particular the most significant contributions. We discuss first the
long-junction limit $\Delta\gg \max[E_T,k_B T,\mu]$, and consider the
effects of a finite $\Delta$ as corrections in the following sections.

In the limit $\Delta \gg \max[E_T,k_B T,\mu]$ the induced potentials
depend only on the temperatures $T_1$ and $T_2$ of the normal
reservoirs. This is due to the fact that the kinetic equations for
$|E| < \Delta$ are decoupled from the distribution function in the
superconductors. In this limit, the spectral currents for $|E| >
\Delta$ are negligible, so they do not affect the integrated
currents. This is because at these energies the distribution function
is nearly constant in the system ($\fL\approx1$, $\fT\approx0$), so
there is no dissipative current. Additionally, at energies $E\gg E_T$,
the coefficient $\jS \approx 0$, so there is no non-dissipative
supercurrent contribution in the current densities.

If we take the limit $\Delta\rightarrow\infty$ in
expression~\eqref{eq:analapproxpot}, approximate $\T\approx0$ and
neglect the energy dependence of $\DT$ completely, we obtain the
dominant term in the induced potentials
\begin{equation}\label{eq:tpapprox}
  \begin{split}
    \mu_{\text{sc},1/2}
    &=
    \frac{1}{2}
    \frac{R_5 (2 R_{4/3} + R_5)}%
         {(R_1 + R_2 + R_5)(R_3 + R_4 + R_5)}
    \; R_{3/4} \; e ( I_S(T_1) - I_S(T_2) )
    \,.
  \end{split}
\end{equation}
Here, $I_S$ is the observable supercurrent
$I_S(T)=A\sigma\integral[0][\infty]{\E}{\jS\fL^0}$ flowing in the
system when the normal reservoirs are all at the same temperature $T$
and there are no potential differences, and
$\fL^0\equiv\tanh(\E/(2k_B{}T))$. This expression states that a major
contribution to the thermopower arises due to the temperature
dependence \cite{dubos01} of the equilibrium supercurrent.

Estimates for the equilibrium supercurrent appearing in
Eq.~\eqref{eq:tpapprox} exist for SNS junctions. One is given in
Ref.~\onlinecite{wilhelm98}:
\begin{equation} \label{eq:equilibriumformula}
  \begin{split}
    e \, \RSNS \, I_S(T)
    &=
    \frac{64 \pi^{3/2}}{\frac{3}{\sqrt{2}} + 2} \,
    (k_B T / E_T)^{3/2} \,
    e^{ -\sqrt{2\pi} \sqrt{k_B T/E_T} } \,
    \sin(\phi)
    \;
    E_T
    \,,
  \end{split}
\end{equation}
where $E_T$ and $\RSNS=R_3+R_4+R_5$ are the Thouless energy and the
resistance corresponding to the link between the
superconductors. Expression~\eqref{eq:equilibriumformula} is valid for
$\ETSNS \ll k_B T \ll \Delta$. Although the formula applies to a
system where there are no extra normal-metal terminals, we can take
them into account by scaling the supercurrent with the factor
presented in Ref.~\onlinecite{heikkila02scdos}:
\begin{equation} \label{eq:scaleterm}
  c_{\text{scale}} \equiv \frac{A_{SNS} \sigma_{SNS}}%
  {A_{SNS}\sigma_{SNS} + \frac{1}{2}(A_1 \sigma_1 + A_2\sigma_2)}
  \,.
\end{equation}
This holds for $k_B T \gg \ETSNS$, $L_1,L_2\gtrsim\LSNS$, and
$L_5=0$. Obviously, in our case $L_5\ne0$, but the results should
still be useful, since the effect due to the central wire is small
as discussed in Sect.~\ref{sec:spectralcoefs} Now, using
Eqs.~(\ref{eq:tpapprox},\ref{eq:equilibriumformula},\ref{eq:scaleterm}),
we obtain an estimate for the thermally induced potentials in the
linear response limit $T_1=T-\Delta{}T/2$, $T_2=T+\Delta{}T/2$:
\begin{equation} \label{eq:mu_ISanalytic}
  \begin{split}
  \mu_{\text{sc},1/2}
  &=
  32\left(\frac{3}{\sqrt{2}}-2\right)\pi^{3/2}
  \:
  c_{\text{scale}}
  \:
  \frac{R_{3/4} R_5 (2R_{4/3} + R_5)}{(R_1 + R_2 + R_5) \RSNS^2}
  \:
  \sin(\phi)
  \\
  &\quad\times
  \left(\sqrt{2\pi} - 3\sqrt{\ET / (k_B T)}\right)(k_B{}T/\ET)
  \:
  e^{-\sqrt{2\pi}\sqrt{k_B{}T/\ET}}
  \;
  k_B\Delta{}T
  \,.
  \end{split}
\end{equation}
For example, in a structure where all the wires are similar, one
would get for $k_B T = 3~\ET$
\begin{equation}
  \QNS \approx 0.04 \frac{k_B}{e} \approx 3~\muVK \,.
\end{equation}
Comparing to numerical results (see Fig.~\ref{fig:approxs}),
estimate~\eqref{eq:mu_ISanalytic} is found to be useful for
$k_B{}T\gtrsim3~\ET$.  However, in the following we prefer to
calculate the integrals with $\jS$ directly from numerical data, which
gives better estimates at temperatures $k_B T \lesssim \ETSNS$.

Taking the coefficient $\T$ into account but still neglecting the
energy dependence of $\DT$ yields the following correction terms to
the potentials:
\begin{equation}
  \begin{split}
  \mu_{\T,1/2} =
  &\mp
  \frac{R_{1/2}}{R_1+R_2+R_5} \,
  \integral[0][\infty]{\E}{\left(\tanhe{1} - \tanhe{2}\right) \TT[1/2]\,}
  \\
  &\mp
  \frac{R_{3/4} R_5}{(R_1+R_2+R_5)\RSNS} \,
  \integral[0][\infty]{\E}{\left(\tanhe{1} - \tanhe{2}\right) \TT[5]\, }
  \,.
  \label{eq:Tcorrection}
  \end{split}
\end{equation}
Here, $\tanhe{k}\equiv\tanh(\E/(2k_B{}T_k))$. The correction is mostly
small compared to the contribution from the supercurrent, but it is
significant in some cases, for example at high temperatures $k_B
T\gtrsim 10~\ETSNS$. However, $\T$ is not known analytically, so the
integrals have to be calculated numerically.

Thus, we have the following approximations for the potentials induced
by a temperature difference:
\begin{subequations}
\label{eq:approxscT}
\begin{align}
  \label{eq:nsapprox}
  \mu_1 &\approx \mu_{\text{sc},1} + \mu_{\T,1}
  \,,\quad
  \mu_2 \approx \mu_{\text{sc},2} + \mu_{\T,2}
  \,,
  \\
  \label{eq:nnapprox}
  \Delta\mu &\equiv \mu_2 - \mu_1 \approx
  (\mu_{\text{sc},2} - \mu_{\text{sc},1})  + (\mu_{\T,2} - \mu_{\T,1})
  \,.
\end{align}
\end{subequations}
For left--right-symmetric structures, $R_1\!=\!R_2$, $R_3\!=\!R_4$,
these approximations yield $\mu_{1/2}\ne0$ and $\Delta\mu = 0$, i.e. a
finite N-S thermopower but a vanishing N-N thermopower. This is due to
the symmetries $\T[1](x) = -\T[2](L_{1,2} - x)$ and $\TT[5]=0$, which
make the correction~\eqref{eq:Tcorrection} the same for both
potentials $\mu_1$ and $\mu_2$

However, there is also a third contribution, which arises from the
energy dependence of $\DT$. This $\DT$-effect has only a small impact
on $\QNS$, but it is noticeable (see Fig.~\ref{fig:ntp}) in $\QNN$ in
a left--right-symmetric structure. A fair estimate for the effect can
be obtained by evaluating the matrix $\vc{g}$ in
Eq.~\eqref{eq:analapproxpot} using numerical results for $\DT$. For a
left--right-symmetric structure, we can approximate the effect more
roughly with
\begin{equation}\label{eq:DTapprox}
  \Delta\mu \approx
  (\mu_{\text{sc},1}+\mu_{\T,1})
  \integral[0][\infty]{\E}{
    \frac{R_1 + \frac{R_3 R_5}{2 R_3 + R_5}}%
     {\RT_1 + \frac{\RT_3 \RT_5}{2 \RT_3 + \RT_5}}
    \left(\frac{\sech^2(\frac{E}{2 k_B T_1})}{2 k_B T_1}
        - \frac{\sech^2(\frac{E}{2 k_B T_2})}{2 k_B T_2} \right)
  }
  \,,
\end{equation}
where $\RT_k = \MT[,k] R_k$ are the energy-dependent spectral
resistances of the wires. Generally this contribution is
non-negligible only for large $|T_2-T_1|$.

\subsection{Quasiequilibrium limit}

In the calculation above, we neglected the inelastic scattering.
However, this assumption is not too restrictive for the
thermopower, as long as a supercurrent may flow through the normal
metal. To see this, let us consider a ``quasiequilibrium'' system,
where inelastic scattering has relaxed the distribution functions
into Fermi functions with a local chemical potential $\mu(x)$ and
temperature $T(x)$. Assuming the induced potentials are again
small, approximating $\DT \approx 1$, $\DL \approx 1$, we can
integrate the kinetic equations over the energy $E$ (after
multiplying the equation for $j_L$ by $E$). Using the fact that
for quasiequilibrium distribution functions, in the limit where
the induced potential differences are small,
\begin{equation}
  \mu(x) = \integral[0][\infty]{\E}{\fT(x)} \,,\qquad
  T(x)  = \frac{\sqrt{6}}{\pi k_B} \left( \integral[0][\infty]{\E}{
    \E \left(1 - \fL(x)\right)} \right)^{\frac{1}{2}}
  \,,
\end{equation}
we get the kinetic equations in wire $i$
\begin{subequations}
  \begin{gather}
    \partial_x (L_i \partial_x \mu(x)+e R_i I_S^i(T(x))+L_i \Tti(x) \partial_x T^2)=0 \,,\\
    \partial_x^2 T^2(x)=0 \,.
  \end{gather}
\end{subequations}
Here
\begin{equation}
  I_S^i(T(x))=\frac{L_i}{2R_i} \integral{\E}{\jS^i(E)f_L(E;T(x))}
\end{equation}
is the supercurrent flowing in wire $i$. In the studied system,
for a constant temperature, $I_S^4=-I_S^3=I_S^5 \equiv I_S(T)$ and
$I_S^1=I_S^2=0$. Moreover, we may include the $\T$-term through
\begin{equation}
  \Tti(x) = \frac{1}{2}\int\!\!\text{d}\E \T(E,x) \partial_{T^2} f_L(E;T(x)) =
  \int\!\!\text{d}\E \frac{- E \T(E,x)}{8 k_B T(x)^3 \cosh^2(E/2k_B T(x))}\,,
\end{equation}
but it does not have a direct physical interpretation. Both the
spectrum $\jS(E)$ of supercurrent-carrying states and the
anomalous coefficient $\T(E)$ may in principle depend on the
magnitude of inelastic scattering, but our aim here is to relate
the measured thermopower to the temperature dependence of the
actual supercurrent flowing in the system (and to the term
$\Tti(x)$). The supercurrent can be probed separately, and thus
for this effect, we do not need to know the exact form of
$\jS^i(E)$.

The boundary and nodal conditions for the solutions $\mu_i(x)$,
$T_i(x)$ are analogous to the general nonequilibrium case: in the
reservoirs, they get the bulk values, except at the
normal-superconducting boundary, say $x=x_{\rm NS}$ we have (as long
as $k_B T \ll \Delta$)
\begin{equation}
  \partial_x T(x)|_{x=x_{\rm NS}}=0 \,.
\end{equation}
Moreover, at the nodes, the functions are continuous, and the charge
and heat currents are conserved. These equations imply that the
temperature $T(x)$ is constant in wires 3 and 4, and the heat current
obeys the Wiedemann-Franz law in the rest of the wires, i.e., $T^2(x)$
is a linear function of position.

The solutions to these (linear) equations can easily be found. For
simplicity, let us first ignore the term(s) $\Tti(x)$. Denote the
position in each wire by $x$, ranging from $x=0$ at the reservoirs
or the left-hand node in wire 5, to $L_i$ at the other end. In
wires 1 and 2, the solutions are
\begin{subequations}
\begin{align}
  \mu&=\mu_1 = {\rm const}\,, &  \quad \mu&=\mu_2 = {\rm const} \,, \\
  T^2(x)&=T_1^2+(T_{w3}^2-T_1^2)\frac{x}{L_1} \,, &
  T^2(x)&=T_2^2+(T_{w4}^2-T_2^2)\frac{x}{L_2} \,.
\end{align}
In wires 3 and 4, we get
\begin{align}
\mu(x)&=\mu_1 \frac{x}{L_3}\,, & \mu(x)&=\mu_2 \frac{x}{L_4} \,, \\
T&=T_{w3}\,, & \quad T&=T_{w4}\,,
\end{align}
respectively. Finally, in wire 5 the solutions are
\begin{align}
  \mu(x)&=\mu_1 + eR_5 c_5 \frac{x}{L_5} + \frac{eR_5}{L_5}
  \integral[0][x]{x'}{I_S(T(x'))} \,, \\
  T^2(x)&=T_{w3}^2+(T_{w4}^2-T_{w3}^2)\frac{x}{L_5} \,.
\end{align}
\end{subequations}
Here $\mu_1$, $\mu_2$, $T_{w3}$, $T_{w4}$, and $c_5$ are constants
that may be determined by requiring continuity and current
conservation at the nodes. We get
\begin{subequations}
\begin{align}
T_{w3}^2&=\frac{(R_2+R_5)T_1^2+R_1 T_2^2}{R_1+R_2+R_5} \,, \qquad
T_{w4}^2=\frac{(R_1+R_5)T_2^2+R_2 T_1^2}{R_1+R_2+R_5} \,, \\
V_1^{\text{sc}}&=\frac{R_3}{R_{SNS}} \left[-R_4 I_S(T_{w4}) +
  (R_4+R_5)I_S(T_{w3}) - R_5
  \int_{T_{w3}^2}^{T_{w4}^2} \frac{I_S(T)\,\text{d}T^2}{T_{w4}^2-T_{w3}^2}\right] \,,\\
V_2^{\text{sc}}&=\frac{R_4}{R_{SNS}} \left[-(R_3+R_5) I_S(T_{w4}) +
  R_3 I_S(T_{w3}) + R_5
  \int_{T_{w3}^2}^{T_{w4}^2} \frac{I_S(T)\,\text{d}T^2}{T_{w4}^2-T_{w3}^2}\right] \,,\\
c_5&=\frac{V_1^{\text{sc}}}{R_3}-I_S(T_{w3})\,, \qquad R_{SNS} \equiv
R_3+R_4+R_5\,.
\end{align}
\end{subequations}
Thus, the induced voltages $V_1^{\text{sc}}=\mu_1/e,
V_2^{\text{sc}}=\mu_2/e$ in the normal-metal reservoirs are determined
from the observable supercurrent $I_S(T)$ between the superconductors,
in the temperature range determined by $T_{w3}$ and $T_{w4}$.

Including the term $\Tti(x)$ is straightforward but leads to long
expressions. Therefore, we present only the resulting  voltages
$V_i$:
\begin{subequations}
\begin{align}
V_1&=V_1^{\text{sc}}+\frac{R_3}{R_{SNS}}\left(T_{w4}^2-T_{w3}^2\right)
\TTti[5]+(T_{w3}^2-T_1^2)\TTti[1]\\
V_2&=V_2^{\text{sc}}+\frac{R_4}{R_{SNS}}\left(T_{w3}^2-T_{w4}^2\right)
\TTti[5]+(T_{w4}^2-T_2^2)\TTti[2],
\end{align}
\end{subequations}
where
\begin{equation}
\TTti[i]\equiv \frac{1}{L_i} \integral[0][L_i]{x}{\Tti[i](x)}
\end{equation}
is the average of the coefficient $\Tti$ in wire $i$.

Let us now take the linear-response limit around the temperature $T_0$, where
\begin{equation}
T_{1/2}^2=T_0^2 \pm \frac{\Delta T^2}{2},
\end{equation}
and
\begin{equation}
I_S(T) \approx I_S(T_0) + \frac{dI_S}{dT^2} \Delta T^2.
\end{equation}
In practice, the requirement for the validity of these formulae is
$\Delta T^2 \ll \min(T_0^2,E_T^2/k_B^2)$. In this case, always
retaining only terms up to the first order in $\Delta T^2$, we get
for the voltage in the left reservoir
\begin{equation}
V_1^{\text{sc}} = \frac{R_5 R_3 (2R_4+R_5)}{2(R_3+R_4+R_5)(R_1+R_2+R_5)}
\frac{dI_S(T)}{dT^2} \Delta T^2.
\end{equation}
This is the same as the linearized form of
Eq.~\eqref{eq:tpapprox}. Similarly, the linearized form of the
$\T$-correction follows that of Eq.~\eqref{eq:Tcorrection}. This
equivalence in the linear regime is due to the fact that the
considered effect is not dependent on the exact shape of the
distribution functions, and thus, in the linear regime, the
relation between the thermopower and the temperature dependent
equilibrium supercurrent is independent of the strength for
inelastic scattering.

\section{GENERAL BEHAVIOR OF THE THERMOPOWER} \label{sec:general}

\begin{figure}\centering
  \includegraphics{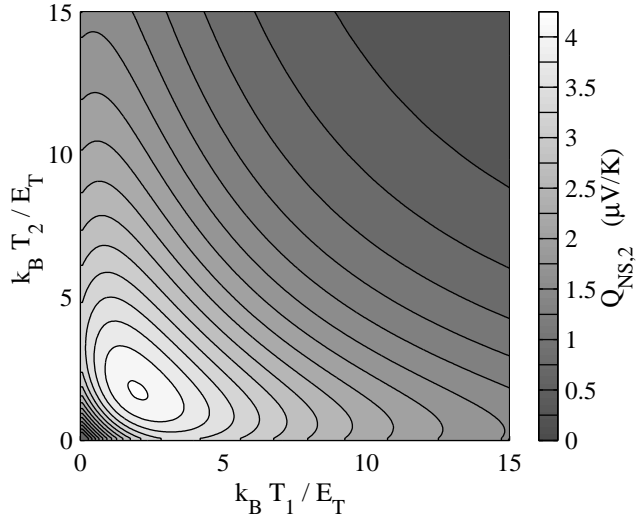}
  \caption{\label{fig:tp} Temperature dependence of the N-S
    thermopower in a left--right-symmetric structure ---
    $L_j,\sigma_j$ and $A_j$ assumed equal. Phase $\phi=\pi/2$ is
    fixed.}
\end{figure}

\begin{figure}\centering
  \subfigure{\raisebox{.05cm}{\includegraphics{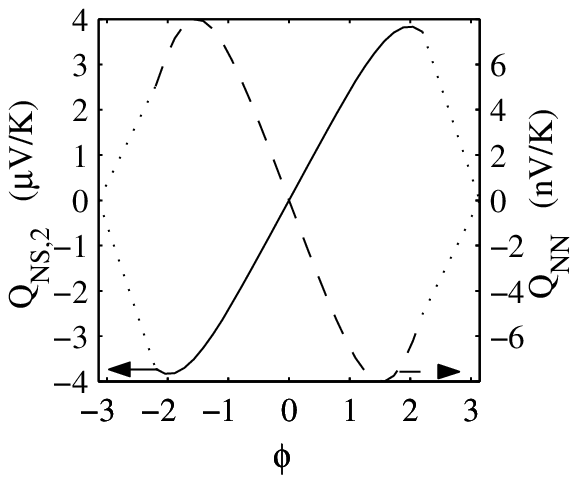}}}
  \subfigure{\includegraphics{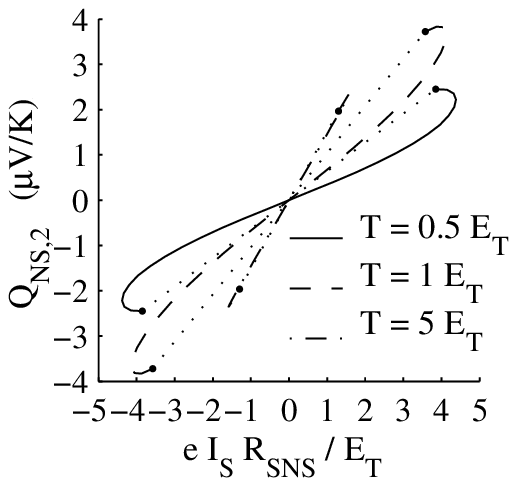}}
  \caption{\label{fig:phase} \label{fig:QIS} Oscillation of the
    thermopower. Left: Phase-oscillation of the thermopower at
    $T_2\approx{}T_1=1~\ET$. Right: Relation between $\QNS$ and the
    non-equilibrium supercurrent $I_S$ at different temperatures
    $T=T_1\approx{}T_2$. Magnitude of the supercurrent here is nearly
    the same as in equilibrium. Due to a numerical convergence
    problem, data for phases $|\phi|>2.1$ is not available, and hence
    substituted with straight line segments (dotted). In the right
    picture, the dotted part also approximately describes the accessible
    non-dissipative regime in the case when the supercurrent is
    externally driven.}
\end{figure}

The typical magnitude and form of the temperature dependence of $\QNS$
is shown in Fig.~\ref{fig:tp}. The figure shows that the magnitude is
$\QNS\sim\muVK$ (often $\QNS[,1]\approx\QNS[,2]$), and that the
relevant energy scale is $\ET$, corresponding to the link between the
superconductors --- which is also the energy scale of the spectral
coefficients.
Moreover, as shown in the preceding sections, the thermopower
oscillates antisymmetrically with the phase difference (see
Fig.~\ref{fig:phase}).

The N-S thermopower shows re-entrant behavior: it vanishes as
$T_1,T_2\rightarrow0$, peaks at an intermediate temperature, and
decays as $T_1,T_2\rightarrow\infty$. The peak near
$T\sim2\ET/k_B$ corresponds to the position where $I_S(T)$ changes
rapidly as a function of the temperature, as
Eqs.~(\ref{eq:tpapprox},\ref{eq:equilibriumformula}) indicate.
Moreover, Eqs.~(\ref{eq:tpapprox},\ref{eq:Tcorrection}) predict
that $\QNS$ decays at least as $T^{-2}$ for $T_1\approx{}T_2$. In
fact, Eq.~\eqref{eq:mu_ISanalytic} implies that the contribution
from the supercurrent should decay exponentially, but the
contribution from $\T$ has a slower speed of decay.

The most significant contribution to $\QNS$ is described by
Eq.~\eqref{eq:tpapprox}: it yields almost all of the effect seen in
Fig.~\ref{fig:tp}. The correction due to $\T$ accounts for the most of
the deviation from the numerical result, and has an effect below 10\%
in relative magnitude for temperatures $T_1,T_2\lesssim10~\ET/k_B$,
but the proportion becomes relatively larger at higher temperatures
--- see Fig.~\ref{fig:approxns}. Moreover, $\DT$ has a negligible
effect on $\QNS$, as seen in Fig.~\ref{fig:approxns}.

The thermopower $\QNS$ with respect to the actual (non-equilibrium)
supercurrent flowing in the structure is shown in Fig.~\ref{fig:QIS}.
For a given temperature, the relation between $\QNS$ and $I_S$ appears
to be linear, in accord with Eq.~\eqref{eq:tpapprox}.

\begin{figure}\centering
  \subfigure[\label{fig:ntp_asym}]{\includegraphics{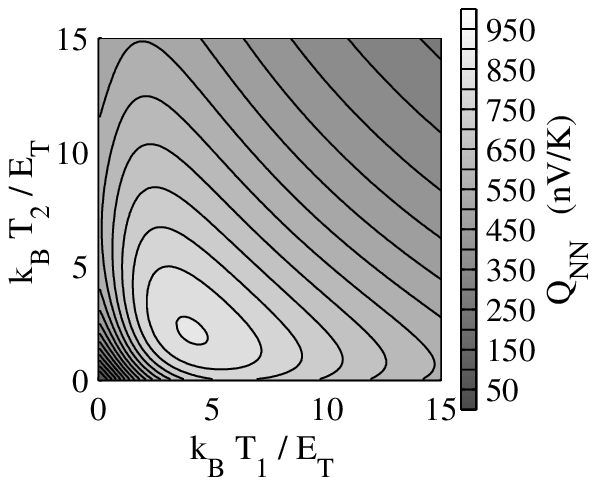}}
  \subfigure[\label{fig:ntp}]{\includegraphics{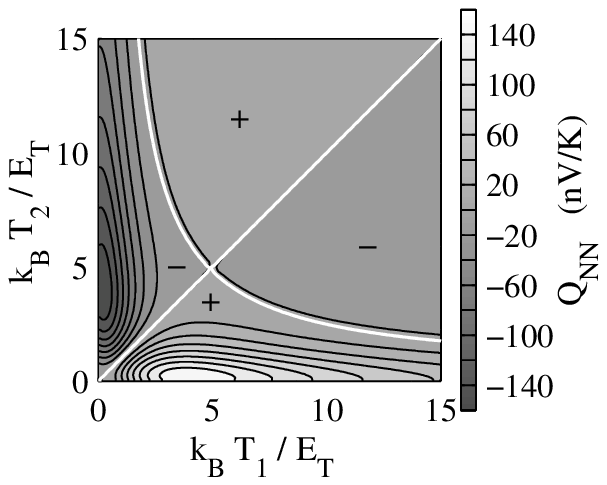}}

  \caption{Temperature dependence of the N-N thermopower. Left: $\QNN$
    in an asymmetric structure, $R_3/2=R_1=R_2=R_4=R_5$ with $A\sigma$
    equal, at $\phi=\pi/2$.  Right: $\QNN$ in a
    left--right-symmetric structure, where the $\DT$-contribution
    dominates. The thick white lines indicate a change of sign for
    $R(T_1)-R(T_2)$, which occurs near the change of sign of $\QNN$.}
\end{figure}

The temperature dependence of $\QNN$ shown in Fig.~\ref{fig:ntp} is
very different from that of $\QNS$, since the structure there is
assumed to be left--right symmetric, so that for $\QNN$ the
contributions from $I_S$ and $\T$ vanish and the $\DT$-effect
dominates (see Fig.~\ref{fig:approxnn}). In this case,
Eq.~\eqref{eq:DTapprox} predicts $\Delta\mu\propto(T_1-T_2)^2$ at
$T_1\approx{}T_2$ in symmetric structures, thus the thermopower
appears only at large temperature differences, i.e. in the non-linear
regime. For asymmetric structures, $\DT$-effect is less important, as
Eq.~\eqref{eq:nnapprox} implies $\Delta\mu\propto(T_1-T_2)$ at
$T_1\approx{}T_2$, and this effect tends generally to wash out the
$\DT$-effect also at large temperature differences. This behavior can
be seen in Fig.~\ref{fig:ntp_asym}, where $\QNN$ has a similar
temperature dependence as $\QNS$ in Fig.~\eqref{fig:tp}. Effects due
to the sample geometry are discussed in more detail in the following
section.

Note that the re-entrant temperature-dependent behavior of the
conductance \cite{charlat96} induced by the proximity effect also
arises due to the energy dependence of $\DT$. Thus, the finite
NN-thermopower in a left--right-symmetric structure may at least
qualitatively be understood to be caused by an induced left--right
asymmetry in the resistances, which makes the potential induced in the
colder electrode different from the one in the hotter
electrode. Indeed, the sign changes in Fig.~\ref{fig:ntp} (along the
diagonal and the $1/T_1$-like curve) occur close to the curves where
the (zero-bias) conductances, \cite{bezuglyi03}
\begin{equation}
  G(T)=\integral[0][\infty]{\E}{
    \tilde{R}^{-1} \; \frac{\sech^2(\E/(2k_B{}T))}{2 k_B T}}
  \,,
  \quad
  \tilde{R} \equiv \MT R
  \,,
\end{equation}
of wires 1 and 2 at the temperatures $T_1,T_2$ become
equal. Approximately such behavior is also expected on the basis of
Eq.~\eqref{eq:DTapprox}.

\begin{figure}
  \subfigure[\label{fig:approxns} ${\QNS[,2]}$ at $T_1\approx{}T_2$]{\includegraphics{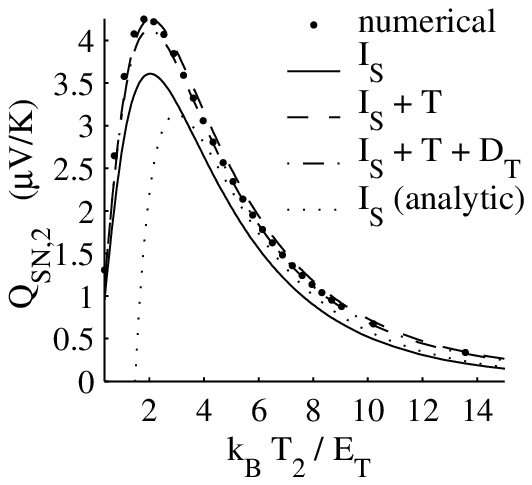}}
  \subfigure[\label{fig:approxnn} $\QNN$ with fixed $T_1=3.6~\ET$]{\includegraphics{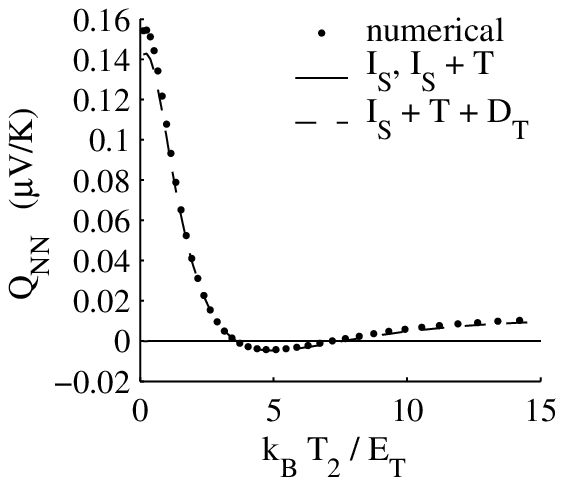}}
  \caption{\label{fig:approxs} Thermopower at $\phi=\pi/2$ in a
    symmetric structure, $L$ and $A\sigma$ assumed equal for each
    wire. The numerical results and different analytic
    approximations from Sect.~\ref{sub:longjunction} are shown. Here,
    ``$I_S$ (analytic)'' is the approximation from
    Eq.~\ref{eq:mu_ISanalytic}.}
\end{figure}

Comparing to numerical results, we find that the approximations
obtained from Eq.~\eqref{eq:analapproxpot} in the long--junction
limit are quantitatively accurate in a symmetric system, both for
$\QNS$ and $\QNN$ (the relative error in $\QNS$ is less than 5\%
nearly everywhere in Fig.~\ref{fig:tp}, although near
$T_1,T_2\approx0$ it is $\sim15\%$, at worst). In setups
geometrically deformed from this, the accuracy may decrease
slightly, but qualitative features are still well retained (in all
cases tested). Equations
~(\ref{eq:analapproxpot},\ref{eq:tpapprox},\ref{eq:Tcorrection},\ref{eq:DTapprox})
may thus be used to provide useful approximations to the induced
potentials, if the spectral coefficients are known.

\section{DEPENDENCE ON SAMPLE GEOMETRY} \label{sec:geometry}

\begin{figure}
  \subfigure[\label{fig:nsasym}${\QNS[,2]}$]{%
    \includegraphics{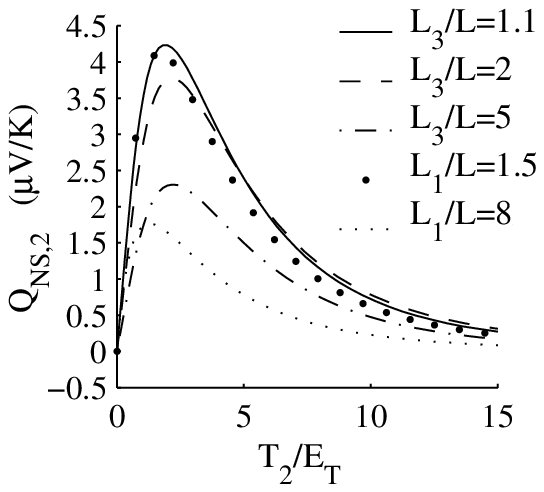}}
  \subfigure[\label{fig:nnasymm} $\QNN$]{\includegraphics{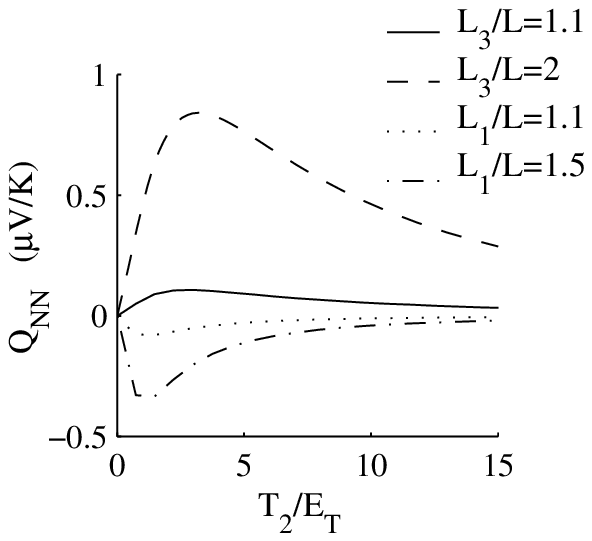}}
  \caption{\label{fig:asymm} Thermopower in an asymmetric structure
    with all except one wire of length $L$, at $T_1\approx{}T_2$,
    $\phi=\pi/2$. $A\sigma$ assumed equal in all wires.}
\end{figure}


The thermopower induced by the proximity effect has a non-trivial
dependence on the exact geometry of the structure: varying the
geometry of the structure changes both the magnitudes and the
temperature dependence of the induced potentials, as seen in
Fig.~\ref{fig:asymm}. The changes arise partly in the kinetic
equations, the effect of which is estimated by the prefactors in the
approximations~(\ref{eq:tpapprox},\ref{eq:Tcorrection}), but also the
spectral equations contribute through the geometry dependence of the
spectral coefficients --- their behavior is discussed in
Sect.~\ref{sec:spectralcoefs}

An effect visible in Fig.~\ref{fig:nsasym} is that compared to the
symmetric structure, the induced potentials typically tend to be
smaller in structures where the lengths of some of the wires are
strongly different from each other.  This behavior appears due to both
the spectral coefficients and the kinetic equations. For example
letting $L_{1,2}/L_{3,4,5}\rightarrow0$ tends to decrease the
thermopower, as then $I_S$ and $\TT[\,]$ are suppressed, while in the
opposite limit most of the temperature drop occurs in wires 1 and 2
where there is little coupling between the charge and energy currents,
leading again to a smaller thermopower (as indicated by the prefactors
in Eqs.~(\ref{eq:tpapprox},\ref{eq:Tcorrection})).

The effect the geometry has on the induced potentials is more
significant for $\QNN$ than for $\QNS$, since the contributions to
$\Delta\mu$ depend strongly on the asymmetry and may also vanish due
to symmetries in the structure. Generally, the approximations imply
that in left--right-asymmetric structures, $\QNN$ should be finite and
roughly resemble the N-S thermopower in temperature dependence, as in
Fig.~\ref{fig:ntp_asym}. Moreover, this thermopower should be
discernible even for small amounts ($\sim10\%$) of asymmetry in
resistances, as can be seen in Fig.~\ref{fig:nnasymm}.

Although the temperature dependence of thermopower generally scales
with $\ETSNS$ similarly to the spectral coefficients, its exact form
depends on the geometry of the structure. That is, the terms due to
$\jS$ and $\TT[\,]$ appearing in the thermopower estimates vary in the
geometry dependence, hence one can alter their relative weights by
adjusting the structure. Moreover, they may also differ in sign and in
the way they depend on the temperature, so their weighing has a direct
effect on the temperature dependence of the thermopower. This is
important especially for the N-N thermopower, which can be seen in
Fig.~\ref{fig:nnasymm} where changes in $L_1$ and $L_3$ cause
qualitatively different results.

\subsection{Thermopower and the central wire} \label{sub:tpwithoutcentralwire}

\begin{figure}\centering
  \includegraphics{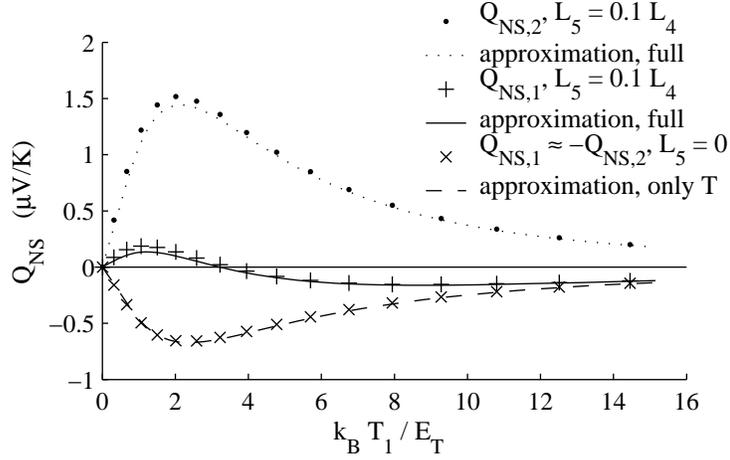}
  \caption{\label{fig:tpL50} N-S thermopower for a structure with the
    central wire short or absent, $L_1=L_2=L_4$ and $A\sigma$ assumed
    equal. Moreover, $T_1\approx{}T_2$ and $\phi=\pi/2$. The ``full
    analytic approximation'' is obtained using
    Eq.~\eqref{eq:analapproxpot} and numerical data for $\DT$, $\jS$
    and $\T$.  The approximation for the contribution due to $\T$
    obtained from Eq.~\eqref{eq:Tcorrection} (dashed) is nearly the
    same in absolute value for all cases shown.}
\end{figure}

As found analytically in Sect.~\ref{sub:longjunction}, the
spectral coefficients that affect the thermopower most are $\jS$
and $\T$. In the long-junction limit, $\jS$ couples the energy and
charge currents in the central wire, and $\TT[1/2]$ do this in the
wires leading to the N-reservoirs. By changing the proportion of
the lengths of these wires, using the factors in
Eqs.~(\ref{eq:tpapprox},\ref{eq:Tcorrection}) as a guide, the
weaker effect due to $\TT[\,]$ may be brought to dominate. In the
following, we consider a special case where $\T$ should manifest:
$L_5\ll{}L_{SNS}$.  (Another possibility suggested by
Eqs.~(\ref{eq:tpapprox},\ref{eq:Tcorrection}) would be
$L_1,L_2\gg{}L_{SNS}$, but in this case, the terms $\TT[1/2]$
become small.)

In the limit $L_5\,/\min[L_1,L_2,L_3,L_4]\rightarrow0$, the
supercurrent-effect approximation~\eqref{eq:tpapprox} yields no
contribution, and the second term in Eq.~\eqref{eq:Tcorrection}
vanishes, leaving only the contribution due to $\TT[1/2]$. If the
structure is left--right-symmetric, then $\T[1/2]=0$, and the
thermopower also vanishes (confirmed by simulations). In fact, this
can be shown exactly by noting that $\fT\equiv0$ is the exact
solution to the kinetic equations in this special case.

In asymmetric structures especially with $L_3\ne{}L_4$, generally
$\TT[1/2]\ne0$ and their sign is the same for both wires 1 and 2.
Hence $\QNS$ becomes finite and there is a large difference
between $\QNS[,1]$ and $\QNS[,2]$, as the contributions from
Eq.~\eqref{eq:Tcorrection} differ in sign for electrodes 1 and 2.
In Fig.~\ref{fig:tpL50} the approximation~\eqref{eq:Tcorrection}
matches the numerical result well: $\TT[1]$ and $\TT[2]$ yield the
dominant contribution.

As the terms~\eqref{eq:tpapprox} due to $I_S$ have the same sign
for both electrodes and vanish as $L_5\rightarrow0$, $\QNS$ may
change its sign with increasing temperature for intermediate
length $L_5$ of the central wire. This is possible because for a
certain range of values for $L_5$, $I_S$ contributes the same
overall amount as $\TT[1/2]$, but in the opposite direction for
either $\QNS[,1]$ or $\QNS[,2]$. As the two contributions have
slightly different temperature dependence, this can result in a
change of sign in the thermopower, for a suitable $L_5$. The
effect is illustrated in Fig.~\ref{fig:tpL50}, where a change of
sign occurs for $L_5=0.1L$. This sign change may partially explain
the results in Ref.~\onlinecite{parsons03}, see
Subs.~\ref{subs:compexp}

\subsection{Effect of decreasing $\Delta/E_T$} \label{sub:finitedelta}

\begin{figure}\centering
  \subfigure[\label{fig:tpdeltans} $\QNS$, $T_0\approx{}T_1\approx{}T_2$]{%
    \includegraphics{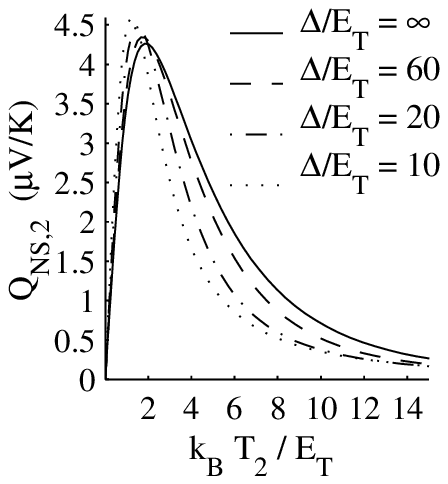}
  }
  \subfigure[\label{fig:tpdeltann} $\QNN$, $T_0=T_1=3.6~\ET$]{%
    \includegraphics{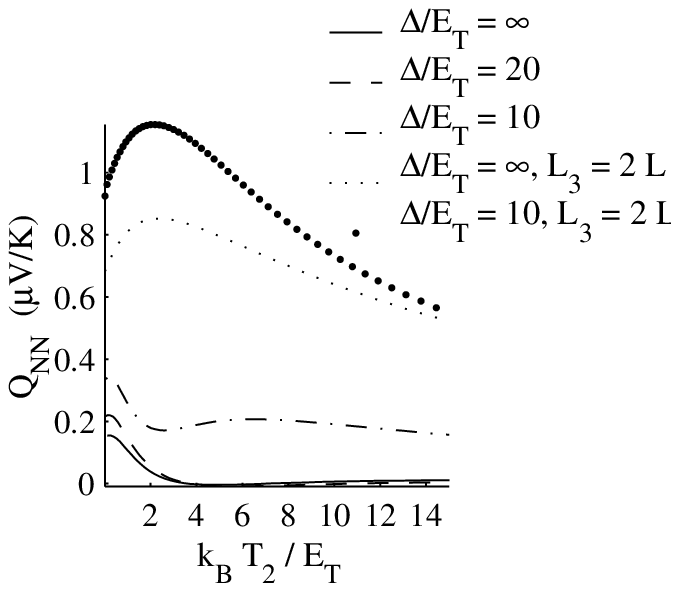}
  }

  \caption{\label{fig:tpdelta} Thermopower for different $\Delta/\ET$.
    Left: $\QNS$ in a structure with all wires equal.  Right:
    $\QNN$ in a structure with $L_1=L_2=L_4=L_5=L$ and $L_3=L$ or
    $L_3=2L$. In both figures, $A\sigma$ is assumed equal for all
    wires. The temperature dependence of $\Delta$ is neglected in the
    figures.}
\end{figure}

In reality, the superconducting energy gap $\Delta$ may be quite
small in systems realized experimentally, and not necessarily much
greater than $\ET$. For example, Eq.~\eqref{eq:E_T} shows that
$\Delta=2$~K corresponds roughly to $10~E_T$ for $L_{SNS} =
1~\mum$ and to $3~E_T$ for $L_{SNS}=0.5~\mum$, for a typical value
$D = 200~\cmss$ of the diffusion constant.

For a finite $\Delta$, the spectral currents from the energies
$E>\Delta$ contribute to the integrated currents. This in turn couples
the temperature $T_0$ of the superconductors to the system, as seen in
Eq.~\eqref{eq:analapproxpot}. Additionally, the spectral coefficients
at energies $E<\Delta$ are modified from their long-junction limits,
and hence the contribution from $E<\Delta$ is also affected.

Estimates for the effect due to the coupling of $T_0$ may be obtained
from Eq.~\eqref{eq:analapproxpot}. Generally, terms proportional to
$(\fL[,0]^0-\fL[,1]^0)$, $(\fL[,0]^0-\fL[,2]^0)$ and
$(\fL[,1]^0-\fL[,2]^0)$ appear for $E>\Delta$, thus any of the
corresponding temperature differences induces potentials. However,
these contributions tend to be smaller than those from $E<\Delta$, at
least for $\Delta/\ET\gtrsim10$ --- supposing of course that $T_1-T_2$
is of the order $T_1-T_0$ or $T_2-T_0$, whichever is larger.

Figure~\ref{fig:tpdeltans} shows the effect of decreasing $\Delta/\ET$
on the N-S thermopower. The energy scale of the temperature dependence
seems to be compressed towards $T=0$, but no significant qualitative
changes occur. The result is similar if $T_1=T_0$ are kept fixed and
$T_2$ is varied.

As $\QNN$ is generally smaller than $\QNS$, the relative changes tend
to be larger in it. This can be seen in Fig.~\ref{fig:tpdeltann},
where the contribution from $E>\Delta$ for $\Delta/\ET=10$ peaks near
$T_2=6~\ET$ in the left--right-symmetric structure, and is also the
dominant effect there. For the asymmetric structure, the change is
relatively much less significant.

\section{COMPARISON WITH RECENT WORK} \label{sec:comparison}

In this section, we compare the theory presented in this paper and
in Ref.~\onlinecite{virtanen04} to recent experiments and other
theories on the thermopower in proximity effect structures.

\subsection{Comparison to experiments}\label{subs:compexp}

To our knowledge, there exist so far five published experiments
from two groups probing the thermopower in Andreev interferometers
\cite{eom98,dikin02,dikin02b,parsons03,parsons03b}. One can
distinguish three features of these experiments to which our
theory can be compared: magnitude of the observed effect and the
dependence on the temperature and magnetic flux.

In Ref.~\onlinecite{eom98}, the temperature gradient produced by the
heating current was not explicitly measured, but only estimated using
a heating model. Such a procedure yielded too large an estimate (of
the order of $\mu$V/K) for the measured NN-thermopower.\cite{dikin02b}
This observation is supported by Refs.~\onlinecite{dikin02,dikin02b},
which discuss further measurements by the same group, now with the
direct measurement of the temperature. Below, we concentrate on the
results of Ref.~\onlinecite{dikin02} --- the observations in
Ref.~\onlinecite{dikin02b} are quite analogous. These measurements
yield a thermopower of the order of 50-60 nV/K at the temperature of
the order of one Thouless temperature corresponding to the distance
between the two superconductors. The measured structure slightly
differed from that considered in our manuscript, as the
superconductors were fabricated directly on top of the normal
wires. However, we may assume the N-S contact resistances to play a
similar role as the resistances of the wires 3 and 4 (see also
Subs.~\ref{subs:theorycomp}). We also note that the magnitude of the
NN thermopower greatly depends on the asymmetry in the structure. For
example, taking the following estimates for the resistances of the
five wires: $R_1=5~\Omega$, $R_2=5.5~\Omega$, $R_3=R_4=0.5~\Omega$,
$R_5=5~\Omega$ (note that mostly only the ratios of the resistances
are relevant) and assuming $E_T/k_B=275$~mK (corresponding to a wire
of length 700~nm) we get $Q_{NN}=60$~nV/K at $T=295$~mK,
$\phi=\pi/2$. This is very close to the experimental results.

Concerning a comparison to the experiments in
Ref.~\onlinecite{parsons03}, we obtain NS thermopower which is larger
by an order of magnitude (of the order of 2 $\mu$V/K, whereas the
experiments report some 70 nV/K). However, it is not totally clear to
us where the normal-metal reservoirs in this experiment reside, and
hence, which values should be used for the resistances $R_1$ and
$R_2$.

Our theory predicts a nonmonotonic thermopower as a function of
the lattice temperature, with a maximum at around $E_T/k_B$, given
by the Thouless energy corresponding to the wire length between
the superconductors. This is in agreement with Fig.~4 in
Ref.~\onlinecite{eom98}. The temperature scale of the dependence
$Q(T)$ in Refs.~\onlinecite{parsons03,parsons03b} is of the order
of $E_T/k_B$, but these papers also report a sign reversal of the
thermopower as a function of $T$. As indicated in
Subs.~\ref{sub:tpwithoutcentralwire}, this may be possible in a
suitable (left--right asymmetric) geometry where the overall
supercurrent and $\T$-term contributions to the thermopower are
comparable. Another possible source for the sign change is the
additional term induced by the coupling of the superconductor
temperature $T_0$ at high temperatures.

As shown in Subs.~\ref{sub:phasedep}, the exact solution of our
equations (within the assumption of electron-hole symmetry) is an
antisymmetric function of the flux (phase) with respect to zero
flux. This is in agreement with the symmetries obtained for the
"Parallelogram" structure in Ref.~\onlinecite{eom98}, and those
measured in
Refs.~\onlinecite{dikin02,dikin02b,parsons03,parsons03b}. However,
our results cannot explain the symmetric flux dependence seen in
the "House" interferometer in Ref.~\onlinecite{eom98}.

\subsection{Comparison with other theories}\label{subs:theorycomp}

The first theoretical discussion of thermoelectric phenomena under
the superconducting proximity effect by Claughton and Lambert
\cite{claughton96} showed how the thermoelectric coefficients can
be calculated from the scattering theory. The general qualitative
features one can extract from the analytic formulae are valid in
any kind of structures composed of normal metals and
superconductors. However, the numerical simulations of these
systems (included, for example, in
Refs.~\onlinecite{claughton96,heikkila00}) suffer from the small
size of the simulated structures which makes it difficult to
differentiate between the effects related with electron-hole
asymmetry (large in the small simulated structures) from the
dominant effects in realistic experimental samples with thousands
of channels.

In Refs.~\onlinecite{seviour00,kogan02}, the thermopower of
normal-superconducting structures is considered in the limit of a
weak proximity effect (with large NS interface resistance) and in
the linear regime. To compare our theory to those in their works,
let us set $T_2=T_0+\delta T/2$, $T_1=T_0-\delta T/2$ in
Eq.~\eqref{eq:tpapprox} for the S-N thermopower in a left--right
symmetric structure,
\begin{equation}\label{eq:differentialQSN}
  Q_{SN} = \frac{\mu_{1,2}}{\delta T} = \frac{1}{4}
  \frac{R_5}{2 R_{1,2} + R_5} \; \frac{1}{(k_B T_0)^2} \;
  \integral[0][\infty]{E}{E \jS' \cosh^{-2}\left( \frac{E}{2 k_B T_0} \right)}
  \,.
\end{equation}
Here, $\jS'=L_3 \jS$. This is analogous to Eq.~(7) in
Ref.~\onlinecite{seviour00} and to the upper line of Eq.~(8) in
Ref.~\onlinecite{kogan02}. The primary coefficient $g_{z+}$ in
these equations is proportional to the spectral supercurrent
across the normal-superconducting interface, analogous to $\jS'$
in Eq.~\eqref{eq:differentialQSN}.

The lower line of Eq.~(8) in Ref.~\onlinecite{kogan02} describes a
thermopower arising between the normal-metal reservoirs. This term
arises from the spectral supercurrent across the interface at the
energies above the superconducting gap $\Delta$ and is thus
similar to the contributions discussed in
Subs.~\ref{sub:finitedelta} As shown there, these contributions
are mostly important only for highly left--right symmetric
structures in the case when $L_{SNS}$ is not much larger than the
superconducting coherence length, or when $T_1$ or $T_2$ are close
to $\Delta/k_B$.

\section{BEHAVIOR OF SPECTRAL COEFFICIENTS} \label{sec:spectralcoefs}

Since the thermopower is induced by the spectral coefficients
$\jS$ and $\TT$, we describe here briefly their dependence on the
energy and on the sample geometry.

\subsection{Spectral supercurrent}

\begin{figure}\centering
  \includegraphics{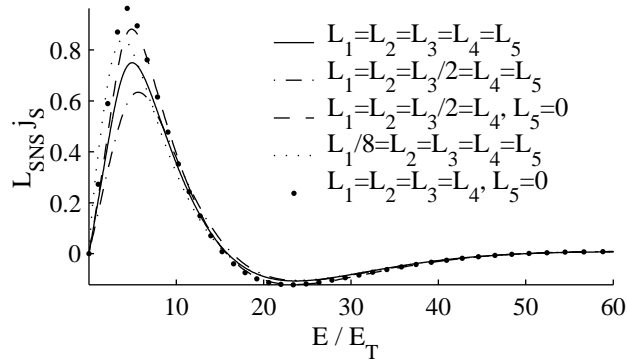}
  \caption{\label{fig:jS_length_ratios}
    The spectral supercurrent $\jS$ for different ratios of
    wire lengths.}
\end{figure}

The spectral supercurrent $\jS$ is the better-known of the two
main contributors for the thermopower. In structures such as
considered here, it typically has the energy dependence shown in
Fig.~\ref{fig:jS_length_ratios}: the characteristic energy scale
is $\ETSNS$, and the dimensionless quantity $L_{SNS}\jS$ does not
depend greatly on the exact geometry of the structure (supposing
$L_1,L_2\gtrsim{}L_{SNS}$). The behavior of $\jS$ is discussed in
detail in Ref.~\onlinecite{heikkila02scdos}, and the results are
also mostly applicable to the structure considered here, as the
central wire typically causes no significant qualitative changes.

\subsection{Anomalous $\T$-term}

\begin{figure}\centering
  \subfigure{\includegraphics{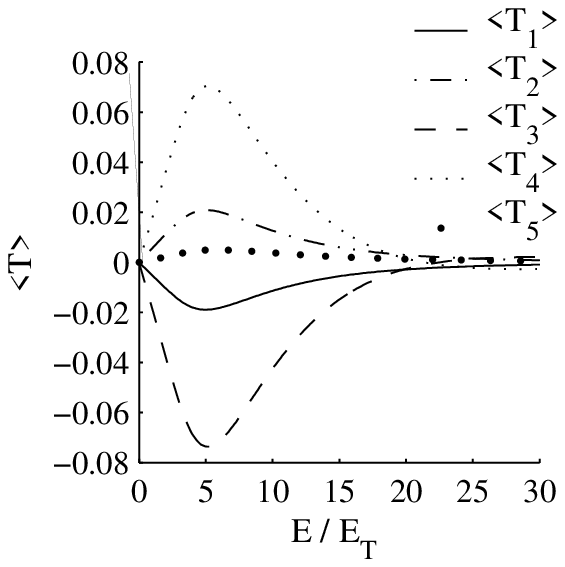}}
  \subfigure{\raisebox{0.1cm}{\includegraphics{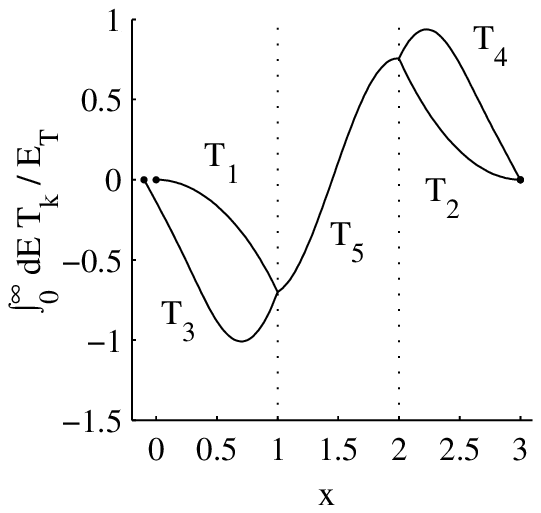}}}
  \caption{\label{fig:T_energy} \label{fig:T_spatial} Behavior of $\T$
    in a structure with $L_1=L_2=L_3/1.1=L_4=L_5$ and $A\sigma$ equal,
    at $\phi=\pi/2$.  Left: The energy-dependence of the averaged
    dimensionless coefficient
    $\TT[\,]=\frac{1}{L_k}\integral[0][L_k]{x}{\T}$ in different
    wires.  Right: Spatial behavior of
    $\integral[0][\infty]{\E/\ET}{\T}$.  Different branches correspond
    to $\T[k]$ in different wires.  Locations of the two nodes are
    indicated by the dotted lines, and reservoir interfaces are
    denoted by dots.}
\end{figure}

The $\T$-term~\eqref{eq:T}, whose effect is neglected in most
previous papers, is a coefficient appearing in the non-equilibrium
part of the supercurrent, and it has no clear physical
interpretation. Nonetheless, it affects the potentials induced due
to a temperature difference, and provides in some cases the
dominant contribution which may be large by itself. Hence it is
interesting to examine the energy and spatial behavior of this
coefficient, which is here considered in the long-junction limit
$\Delta \gg \ET$.

The typical energy dependence of the average $\T$ in different
wires of the structure in Fig.~\ref{fig:kinetic-wires} is shown in
Fig.~\ref{fig:T_energy}. Its magnitude is approximately less than
10\% of the spectral supercurrent $L_{SNS}\jS$, but the energy
dependence is rather similar: a single peak appears on the scale
of $\ETSNS$ although there is less oscillation than in
$L_{SNS}\jS$.

The spatial dependence shown in Fig.~\ref{fig:T_spatial} gives
insight to the magnitudes of $\T$ in different wires. Here, $\T$
vanishes at the reservoirs and changes its sign near the center of
the SNS link, forming peaks between the S-reservoirs and the
center. This leads to $\TT[5]\approx0$ in nearly
left--right-symmetric structures, as seen in
Fig.~\ref{fig:T_energy}. Moreover, $\T[3]$ and $\T[4]$ are
typically the largest and the two differ in sign.

In the wires 1 and 2 leading to the N-reservoirs, $\T$ decays
monotonically to zero on the length scale of $L_{SNS}$ (according to
numerical results). In fact, no sign changes are possible for
$\T[1/2]$, as $\chi\equiv\const{x}$ in these wires. Moreover, note
that for the quantities $\TT[1/2]$ relevant for the thermopower, this
kind of a decay would imply that $\TT[1/2]$ vanish for
$L_{1,2}/L_{SNS}\rightarrow\infty$.

The magnitude and the sign of $\T$ in the wires 1 and 2 depend on
the points where the wires are connected to the SNS link, since
$\T$ is continuous in the structure. Moreover, if the N-reservoirs
are far ($L_1,L_2\gtrsim{}L_{SNS}$), the choice of the connection
points does not affect much the $\T$ in the SNS link, hence one
may estimate the qualitative behavior of $\T$ in wires 1 and 2 by
looking at Fig.~\ref{fig:T_spatial}.

Using the symmetry arguments from Subs.~\ref{sub:phasedep}, one
finds that $\T$ oscillates antisymmetrically with the phase
difference. Moreover, the overall phase behavior of $\T$ seems to
be similar to that of $\jS$, according to numerical results.

\section{DISCUSSION} \label{sec:discussion}

In this paper, we have shown how the presence of the supercurrent
can lead to a finite thermopower in an Andreev interferometer even
in the presence of complete electron-hole symmetry. Generally, the
thermopower probes the temperature dependence of the supercurrent,
which in the long-junction limit is determined by the Thouless
energy corresponding to the distance between the superconductors.
The actual magnitude of the thermopower also strongly depends on
the studied geometry of the sample, which has to be taken into
account when comparing to the experiments.

Another interesting result is the thermopower induced by the
"anomalous" coefficient $\T$. This coefficient has been identified
in a number of theoretical publications (for example,
Refs.~\onlinecite{seviour00,kogan02,heikkila03,yip98,lee99,bezuglyi03b}
--- it is frequently referred to as the "anomalous current" as its
properties are not very well known) but typically its effect is
neglected and we are not aware of any other observable whose
behavior would be dictated by $\T$.

The theory presented in this paper and in Ref.~\onlinecite{virtanen04}
seems to be in fair agreement with most of the experimental results
published so far (see Subs. \ref{subs:compexp}). However, unless
device geometry (resistances of the wires) are quite well known, it is
difficult to make a detailed comparison of the magnitude and
temperature dependence of the measured thermopower. There exists also
one experimental result (the symmetric oscillations of the thermopower
in the "House" interferometer in Ref.~\onlinecite{eom98}) which
clearly cannot be explained with the present theory.

Confirming the theory clearly calls for more experiments on the
subject. Rather than as a function of the applied flux, one could
probe the thermopower by externally applying a supercurrent between
the superconductors (c.f., Fig.~\ref{fig:QIS}). In this way, one would
also be able to compare the observed thermopower to the temperature
dependence of the equilibrium (critical) supercurrent, essentially
with no fitting parameters --- the normal-state resistances can be
fairly accurately measured in the considered multiprobe
structure. Further, it would be interesting to confirm the relation
between the thermopower and the "anomalous" term $\T$, either by
substracting the supercurrent part from the results or, for example,
in a structure without the central wire. The effect of the additional
terms, present in short wires (due to a finite $\Delta/E_T$) or highly
symmetric structures (due to the proximity-induced temperature
dependence in the resistances), would also be worth studying, but
perhaps more difficult to realize in practice.

\section*{ACKNOWLEDGMENTS}
We thank Frank Hekking, Nikolai Kopnin, Mikko Paalanen and Jukka
Pekola for useful discussions.



\begin{thebibliography}{10}

\bibitem{ginzburgnobellecture}
V.~L. Ginzburg.
Nobel lecture (2003).
\\http://www.nobel.se/physics/laureates/2003/ginzburg-lecture.html.

\bibitem{ginzburg44}
V.~L. Ginzburg.
{\em Zh. \'Eksp. Teor. Fiz.\/}, {\bf 14}, 177 (1944).
[J. Phys. USSR 8, 148 (1944)].

\bibitem{ashcroft}
N.~W. Ashcroft and N.~D. Mermin.
{\em Solid-state physics\/}.
Saunders College Publishing (1967).
Chapters 12--13.

\bibitem{vanharlingen82}
D.~J. van Harlingen.
{\em Physica\/}, {\bf 109 \& 110B}, 1710 (1982).

\bibitem{galperin02}
Y.~M. Galperin, V.~L. Gurevich, V.~I. Kozub, and A.~L. Shelankov.
{\em Phys. Rev. B\/}, {\bf 65}, 064531 (2002).

\bibitem{pethick79}
C.~J. Pethick and H.~Smith.
{\em Phys. Rev. Lett.\/}, {\bf 43}, 640 (1979).

\bibitem{clarke79}
J.~Clarke, B.~R. Fjordb{\o}ge, and P.~E. Lindelof.
{\em Phys. Rev. Lett.\/}, {\bf 43}, 642 (1979).

\bibitem{schmid79}
A.~Schmid and G.~Schön.
{\em Phys. Rev. Lett.\/}, {\bf 43}, 793 (1979).

\bibitem{clarke80}
J.~Clarke and M.~Tinkham.
{\em Phys. Rev. Lett.\/}, {\bf 44}, 106 (1980).

\bibitem{raimondilambert}
C.~J. Lambert and R.~Raimondi.
{\em J. Phys.: Condens. Matter\/}, {\bf 10}, 901 (1998).

\bibitem{courtoispannetier}
H.~Courtois and B.~Pannetier.
{\em J. Low Temp. Phys.\/}, {\bf 118}, 599 (2000).

\bibitem{belzig99}
W.~Belzig, F.~K. Wilhelm, C.~Bruder, G.~Sch\"on, and A.~D. Zaikin.
{\em Superlatt. Microstruct.\/}, {\bf 25}, 1251 (1999).

\bibitem{dubos01}
P.~Dubos, H.~Courtois, B.~Pannetier, F.~K. Wilhelm, A.~D. Zaikin, and
  G.~Sch\"on.
{\em Phys. Rev. B\/}, {\bf 63}, 064502 (2001).

\bibitem{charlat96}
P.~Charlat, H.~Courtois, {P}h. Gandit, D.~Mailly, A.~F. Volkov, and
  B.~Pannetier.
{\em Phys. Rev. Lett.\/}, {\bf 77}, 4950 (1996).

\bibitem{stoof96}
{Y}u. V.~Nazarov and T.~H. Stoof.
{\em Phys. Rev. Lett.\/}, {\bf 76}, 823 (1996).

\bibitem{bezuglyi03}
E.~V. Bezuglyi and V.~Vinokur.
{\em Phys. Rev. Lett.\/}, {\bf 91}, 137002 (2003).

\bibitem{claughton96}
N.~R. Claughton and C.~J. Lambert.
{\em Phys. Rev. B\/}, {\bf 53}, 6605 (1996).

\bibitem{eom98}
J.~Eom, C.-J. Chien, and V.~Chandrasekhar.
{\em Phys. Rev. Lett.\/}, {\bf 81}, 437 (1998).

\bibitem{dikin02}
D.~A. Dikin, S.~Jung, and V.~Chandrasekhar.
{\em Phys. Rev. B\/}, {\bf 65}, 012511 (2002).

\bibitem{dikin02b}
D.~A. Dikin, S.~Jung, and V.~Chandrasekhar.
{\em Europhys. Lett.\/}, {\bf 57}, 564 (2002).

\bibitem{parsons03}
A.~Parsons, I.~A. Sosnin, and V.~T. Petrashov.
{\em Phys. Rev. B\/}, {\bf 67}, 140502 (2003).

\bibitem{parsons03b}
A.~Parsons, I.~A. Sosnin, and V.~T. Petrashov.
{\em Physica E\/}, {\bf 18}, 316 (2003).

\bibitem{heikkila00}
T.~T. Heikkil\"a, M.~P. Stenberg, M.~M. Salomaa, and C.~J. Lambert.
{\em Physica B\/}, {\bf 284-8}, 1862 (2000).

\bibitem{seviour00}
R.~Seviour and A.~F. Volkov.
{\em Phys. Rev. B\/}, {\bf 62}, 6116 (2000).

\bibitem{kogan02}
V.~R. Kogan, V.~V. Pavlovskii, and A.~F. Volkov.
{\em Europhys. Lett.\/}, {\bf 59}, 875--881 (2002).

\bibitem{virtanen04}
P.~Virtanen and T.~T. Heikkil\"a. {\em Phys. Rev. Lett.} {\bf 92},
177004 (2004).

\bibitem{heikkila03}
T.~T. Heikkil\"a, T.~V\"ansk\"a, and F.~K. Wilhelm.
{\em Phys. Rev. B\/}, {\bf 67}, 100502(R) (2003).

\bibitem{nazarov99}
Y.~V. Nazarov.
{\em Superlatt. Microstruct.\/}, {\bf 25}, 1221 (1999).

\bibitem{rammer86}
J.~Rammer and H.~Smith.
{\em Rev. Mod. Phys.\/}, {\bf 58}, 323 (1986).

\bibitem{usadel70}
K.~D. Usadel.
{\em Phys. Rev. Lett.\/}, {\bf 25}, 507 (1970).

\bibitem{heikkila02scdos}
T.~T. Heikkil\"a, J.~S\"arkk\"a, and F.~K. Wilhelm.
{\em Phys. Rev. B\/}, {\bf 66}, 184513 (2002).

\bibitem{hansch84}
W.~H\"ansch.
{\em Phys. Rev. B\/}, {\bf 31}, 3504 (1984).

\bibitem{wilhelm98}
F.~K. Wilhelm, G.~Schön, and A.~D. Zaikin.
{\em Phys. Rev. Lett.\/}, {\bf 81}, 1682 (1998).

\bibitem{yip98}
S.-K. Yip.
{\em Phys. Rev. B\/}, {\bf 58}, 5803 (1998).

\bibitem{lee99}
S.-W. Lee, A.~V. Galaktionov, and C.-M. Ryu.
{\em J. Korean Phys. Soc.\/}, {\bf 34}, S193 (1999).

\bibitem{bezuglyi03b}
E.~V. Bezuglyi, V.~S. Shumeiko, and G.~Wendin.
{\em Phys. Rev. B\/}, {\bf 68}, 134506 (2003).

\end{thebibliography}
\end{document}